\documentclass[twocolumn]{aastex63}

\usepackage{longtable}
\usepackage{graphicx}
\usepackage{graphics}
\usepackage{color}
\usepackage{epsfig}
\usepackage{rotating}
\usepackage{subfigure}
\usepackage{float}
\usepackage{url}
\usepackage{hyperref}
\usepackage{breakurl}
\usepackage{scalerel}
\usepackage[titletoc,title]{appendix}
\usepackage{relsize}
\usepackage{bm}
\usepackage{natbib}

\shortauthors{Y. Jiang et al.}
\shorttitle{IRAM 30m CO observations of NGC~4594}

\begin{document}

\title{CO-CHANGES I: IRAM 30m CO Observations of Molecular Gas in the Sombrero Galaxy}

\author{Yan Jiang}
\affiliation{Purple Mountain Observatory, Chinese Academy of Sciences, 10 Yuanhua Road, Nanjing 210023, People’s Republic of China}
\affiliation{School of Astronomy and Space Sciences, University of Science and Technology of China, Hefei 230026, China}

\author[0000-0001-6239-3821]{Jiang-Tao Li}
\affiliation{Purple Mountain Observatory, Chinese Academy of Sciences, 10 Yuanhua Road, Nanjing 210023, People’s Republic of China}

\author{Yu Gao}
\affiliation{Department of Astronomy, College of Physical Science and Technology, Xiamen University, Xiamen 361005, China}
\affiliation{Purple Mountain Observatory/Key Lab of Radio Astronomy, Chinese Academy of Sciences, Nanjing 210034, China}

\author{Joel N. Bregman}
\affiliation{Department of Astronomy, University of Michigan, 311 West Hall, 1085 S. University Ave, Ann Arbor, MI, 48109-1107, U.S.A.}

\author{Li Ji}
\affiliation{Purple Mountain Observatory, Chinese Academy of Sciences, 10 Yuanhua Road, Nanjing 210023, People’s Republic of China}

\author{Xuejian Jiang}
\affiliation{Research Center for Intelligent Computing Platforms, Zhejiang Laboratory, Hangzhou 311100, China}

\author{Qinghua Tan}
\affiliation{Purple Mountain Observatory, Chinese Academy of Sciences, 10 Yuanhua Road, Nanjing 210023, People’s Republic of China}

\author{Jianfa Wang}
\affiliation{Purple Mountain Observatory, Chinese Academy of Sciences, 10 Yuanhua Road, Nanjing 210023, People’s Republic of China}
\affiliation{School of Astronomy and Space Sciences, University of Science and Technology of China, Hefei 230026, China}

\author[0000-0002-9279-4041]{Q. Daniel Wang}
\affiliation{Department of Astronomy, University of Massachusetts, Amherst, MA 01003, U.S.A.}

\author{Yang Yang}
\affiliation{Purple Mountain Observatory, Chinese Academy of Sciences, 10 Yuanhua Road, Nanjing 210023, People’s Republic of China}

\correspondingauthor{Jiang-Tao Li}
\email{pandataotao@gmail.com}


\nonumber

\begin{abstract}
Molecular gas plays a critical role in explaining the quiescence of star formation (SF) in massive isolated spiral galaxies, which could be a result of either the low molecular gas content and/or the low SF efficiency. We present IRAM 30m observations of the CO lines in the Sombrero galaxy (NGC~4594), the most massive spiral at $d\lesssim30\rm~Mpc$. We detect at least one of the three CO lines covered by our observations in all 13 observed positions located at the galactic nucleus and along a $\sim25\rm~kpc$-diameter dusty ring. The total extrapolated molecular gas mass of the galaxy is $M_{\rm H_2}\approx4\times10^{8}\rm~M_\odot$. The measured maximum CO gas rotation velocity of $\approx379\rm~km~s^{-1}$ suggests that NGC~4594 locates in a dark matter halo with a mass $M_{\rm200}\gtrsim10^{13}\rm~M_\odot$. Comparing to other galaxy samples, NGC~4594 is extremely gas poor and SF inactive, but the SF efficiency is apparently not inconsistent with that predicted by the Kennicutt-Schmidt law, so there is no evidence of enhanced SF quenching in this extremely massive spiral with a huge bulge. We also calculate the predicted gas supply rate from various sources to replenish the cold gas consumed in SF, and find that the galaxy must experienced a starburst stage at high redshift, then the leftover or recycled gas provides SF fuels to maintain the gradual growth of the galactic disk at a gentle rate.
\end{abstract}

\section{Introduction}\label{sec:Introduction}

Molecular gas plays a critical role in star formation (SF) and galaxy evolution. It is a key point in the chain of gas recycling in and out of galaxies, often forms via the cooling and accretion of the circum-galactic and/or inter-galactic medium (CGM and/or IGM), then condenses and forms stars, which further feedback metal-enriched materials back into the CGM and IGM (e.g., \citealt{Kennicutt12} and references therein). As the direct SF material, spatially resolved study of the molecular gas together with the SF properties in nearby galaxies gives us direct insights on the SF processes and efficiency (e.g., \citealt{Young91,Kennicutt98,Crocker12}).

Over the past few decades, molecular gas in nearby galaxies have been studied via various tracers, typically rotational or vibrational emission lines in millimeter-wave (mm-wave) or near-IR (e.g., \citealt{Young91,Kennicutt03,Kennicutt12,Gao04,Veilleux09,Wu10}). The CO lines, in particular the low-$J$ (e.g., $J=1-0$) transitions in mm-wave, are the most commonly adopted molecular gas tracers, because they are abundant and easily excited. In addition to the most abundant $^{12}\rm C$ and $^{16}\rm O$, the CO molecule could also be made of some other carbon and oxygen isotopes. One of the most important such isotopologues is $^{13}$CO, which emits at a similar rest frame frequency as $^{12}$CO, but has an intensity strongly affected by the abundance of $^{13}\rm C$ and the optical depth of the molecular cloud (e.g., \citealt{Jim17,Cormier2018}). In comparison to the CO $J=1-0$ transitions, the relative strength of the higher CO transitions (e.g., $J=2-1$ and above) is often affected by the temperature and optical depth of the molecular cloud. The high high-$J$/low-$J$-ratio gas trace the dense molecular clumps with a steep density gradient and a thin CO emitting envelope, or a high gas temperature, which are often found in SF regions with intense heating sources (e.g., \citealt{Hasegawa97,Penaloza17,Leroy22}). Observing different CO lines helps us to more reliably measure the properties (temperature, density, total mass, etc.) of the SF fuel in galaxies. Some scientific questions remain to be explored with systematic CO observations of nearby galaxies, such as: Is molecular gas produced internally and/or externally? How does it form stars under different circumstance, or under what circumstance may SF be quenched?

We herein introduce our IRAM 30m survey of the CO emission lines from the CHANG-ES galaxies (the CO-CHANGES project). The CHANG-ES (Continuum HAlos in Nearby Galaxies - an Evla Survey) project is mainly based on systematic VLA observations of 35 nearby edge-on disk galaxies, enabling the study of cosmic ray (CR) and magnetic field in galactic halos (e.g., \citealt{Irwin12a,Irwin12b,Irwin19a,Wiegert15,Krause18,Krause20}). The survey has led to many follow-up projects, including those based on both the original VLA data (studying the \ion{H}{1} 21-cm line; \citealt{Zheng22a,Zheng22b}), follow-up VLA observations in other bands or configurations (e.g., \citealt{Irwin19b}), or multi-wavelength observations in X-ray (e.g., \citealt{Li13a,Li13b,Li14}), H$\alpha$ (e.g., \citealt{Vargas19,Lu23}), infrared (IR) (e.g., \citealt{Vargas18}), low-frequency radio continuum (e.g., \citealt{Heald21,Stein22}), as well as studies of important point-like sources (e.g., \citealt{Irwin17,Perlman21}). Our IRAM 30m observations in the CO-CHANGES project simultaneously covers the $^{12}$CO~$J=1-0$, $^{13}$CO~$J=1-0$ and $^{12}$CO~$J=2-1$ emission lines from multiple selected positions along the galactic disk of 24 of the 35 CHANG-ES galaxies, with a total exposure time of $\approx81\rm~hrs$ taken from October 2018 to January 2019.

In the present paper, we report an initial case study of the IRAM 30m observations of NGC~4594, a nearby edge-on spiral galaxy also known as the Sombrero galaxy or M104 ($d=12.7\rm~Mpc$, $1^{\prime\prime}=61.7\rm~pc$, inclination angle $i\approx79^{\circ}$; Table~\ref{table:NGC4594}). NGC~4594 is the most massive spiral galaxy within a distance of $d\sim30\rm~Mpc$ \citep{Jarrett19}. It is an isolated Sa galaxy with a large stellar bulge and a low environmental galaxy density (e.g., \citealt{Irwin12a,Li16}). Its central supermassive black hole has a mass of $6.6\times10^8\rm~{M_\odot}$ \citep{Jarrett11}, but appears as a low-luminosity active galactic nucleus (AGN), although a 10-kpc scale radio jet has been discovered in a recent observation \citep{Yang23}. The SF rate is also low for such a massive galaxy (${\rm SFR}\approx0.4\rm~M_\odot~yr^{-1}$, stellar mass $M_*\approx2.6\times10^{11}\rm~M_\odot$ or 4-5 times of the stellar content of the Milky Way (MW); \citealt{Vargas19,Li16,Ogle16,Li17}). This galaxy was observed in CO with the IRAM 30m at only two positions \citep{Bajaja91} and with ALMA covering the dust ring but only at the $^{12}$CO $J=1-0$ line \citep{Sutter22}. We herein present a complete view of multiple CO lines across the dust ring of NGC~4594 (Fig.~\ref{fig:IRAMbeam}), which is of particular interest in understanding the SF properties of such a massive isolated spiral galaxy.

The present paper is organized as follows: we will introduce the observations and data reductions in \S\ref{sec:datareduction}, then present our main results in \S\ref{sec:results}. We will further compare the molecular gas properties of NGC~4594 to other galaxy samples in \S\ref{sec:discussion} and discuss the scientific implications. Our main results and conclusions will be summarized in \S\ref{sec:summary}. Throughout the paper, we quote a 3~$\sigma$ upper limit for the non-detection of the CO lines at some positions, while adopt 1~$\sigma$ errors in any solid detections.


\begin{table}
\vspace{-0.in}
\begin{center}
\caption{Parameters of NGC~4594} 
\footnotesize
\tabcolsep=2.5pt
\begin{tabular}{lcccccccccccccc}
\hline\hline
Parameters & Value & Reference \\
\hline
Type & SA(s)a;LINER;Sy1.9 & 1 \\
Type code & $1.1\pm 0.4$ & 2\\
Distance & 12.7 Mpc & 3 \\
Redshift & 0.003416 & 1 \\
Angular scale & $1^{\prime\prime}=0.0617\rm~kpc$ & 3 \\
Inclination & $79^\circ$ & 3 \\
$v_{\rm rot}$ & $408.0\rm~km~s^{-1}$ & 4 \\
$M_*$ & $(2.6\pm0.4)\times10^{11}\rm~M_\odot$ & 4 \\
$M_{\rm HI}$ & $6.1\times10^8\rm~M_\odot$ & 5 \\
SFR & $(0.43\pm0.04)\rm~M_\odot~yr^{-1}$ & 6 \\
$\Sigma_{\rm SFR}$ & $(1.14\pm0.01)\times10^{-3}\rm~M_\odot~yr^{-1}~kpc^{-2}$ & 6\\
$M_{\rm hot}$ & $5.27\times10^{8}\rm~M_\odot~yr^{-1}$ & 7\\ 
$\dot{M}_{\rm hot}$ & $0.057\rm~M_\odot~yr^{-1}$ & 7 \\ 
\hline
$M_{\rm H_2,^{12}CO}$ & $(4.0\pm0.1)\times10^8\rm~M_\odot$ & 8 \\
$M_{\rm H_2,^{13}CO}$ & $2.9^{+0.3}_{-0.2}\times10^8\rm~M_\odot$ & 8 \\
$\left \langle{\rm \mathsmaller{\mathsmaller{\frac{^{12}CO_{\rm 10}}{^{13}CO_{\rm 10}}}}}\right \rangle$ & $9.2^{+3.3}_{-2.1}$ & 8 \\
$\left \langle{\rm \mathsmaller{\mathsmaller{\frac{^{12}CO_{\rm 21}}{^{12}CO_{\rm 10}}}}}\right \rangle$ & $0.34^{+0.05}_{-0.04}$ & 8 \\
$\left \langle{\tau}_{\rm ^{13}CO}\right \rangle$ & $0.12^{+0.05}_{-0.03}$ & 8 \\
$\left \langle{T}_{\rm K}\right \rangle$ & $41^{+20}_{-7}\rm~K$ & 8 \\
$\Sigma_{\rm H_2}$ & $1.07\pm0.03\rm~M_\odot~pc^{-2}$ & 8 \\
$\Sigma_{\rm H_2+HI}$ & $2.70\pm0.03\rm~M_\odot~pc^{-2}$ & 8 \\
$V_{\rm max,corr}$ & $381\pm32\rm~km~s^{-1}$ & 8 \\
\hline\hline
\end{tabular}\label{table:NGC4594}
\end{center}
The parameters above and below the horizontal line are quoted from the literature and our own measurements in the present paper, respectively. $v_{\rm rot}$ is the inclination corrected rotation velocity. $M_*$ is the stellar mass estimated from the K-band magnitude. $M_{\rm HI}$ is the mass of atomic gas obtained from \citet{Bajaja84} and rescaled to our adopted distance of $12.7\rm~Mpc$. SFR and $\Sigma_{\rm SFR}$ are the star formation rate and the SFR surface density obtained by combining H$\alpha$ and WISE $22\rm~\mu m$ data. $M_{\rm hot}$ and $\dot{M}_{\rm cool}$ are the total mass and radiative cooling rate of the X-ray emitting hot gas. $M_{\rm H_2}$ represents the total mass of the molecular gas measured with the $^{12}$CO~$J=1-0$ and $^{13}$CO~$J=1-0$ lines, respectively. $\left \langle{\rm \mathsmaller{\mathsmaller{\frac{^{12}CO_{\rm 10}}{^{13}CO_{\rm 10}}}}}\right \rangle$, $\left \langle{\rm \mathsmaller{\mathsmaller{\frac{^{12}CO_{\rm 21}}{^{12}CO_{\rm 10}}}}}\right \rangle$, $\left \langle{\tau}_{\rm ^{13}CO}\right \rangle$, and $\left \langle{T}_{\rm K}\right \rangle$ are the median values of the line ratios, the optical depth, and the kinetic temperature of the molecular gas on the dust ring of NGC~4594 (excluding the nuclear region), respectively. $\Sigma_{\rm H_2}$ and $\Sigma_{\rm H_2+HI}$ are the surface densities of the molecular gas and the total molecular+atomic gas, adopting the area of the dust ring calculated in \S\ref{subsec:SFR}. $V_{\rm max,corr}$ is the inclination corrected maximum rotation velocity measured by fitting the position-velocity diagram.\\

References in this table:\\
1. NED (\url{https://ned.ipac.caltech.edu})\\
2. HyperLeda (\url{http://leda.univ-lyon1.fr/})\\
3. \citet{Irwin12a}\\
4. \citet{Li16}\\
5. \citet{Bajaja84}\\
6. \citet{Vargas19}\\
7. \citet{Li13a}\\
8. This paper
\\
\end{table}

\section{Observations and Data Reduction}\label{sec:datareduction}

\subsection{IRAM 30m observations}\label{subsec:Observations}

We observed NGC~4594 with the IRAM 30m telescope in the 2018B semester during two observation runs in October 2018 and January 2019, respectively (Project ID 063-18 and 189-18; PI: Jiang-Tao Li). The observations were taken with the Eight MIxer Receiver (EMIR) in position switching (PSW) mode with the SIS receivers – EMIR E90/E230 \citep{Carter12}, which simultaneously cover the $^{12}$CO $J=1-0$, $J=2-1$, and $^{13}$CO $J=1-0$ lines at rest-frame frequencies of $\nu_{\rm rest,^{12}CO1-0}=115.271\rm~GHz$, $\nu_{\rm rest,^{12}CO2-1}=230.538\rm~GHz$, and $\nu_{\rm rest,^{13}CO1-0}=110.201\rm~GHz$, respectively. We used the fast Fourier Transform Spectrometer (FTS) backend, which provides a frequency resolution of $200\rm~kHz$. The half power beam width (HPBW) is $\approx21.4^{\prime\prime}$ and $\approx10.7^{\prime\prime}$ at 115~GHz and 230~GHz, respectively. 

\begin{figure}
\begin{center}
\epsfig{figure=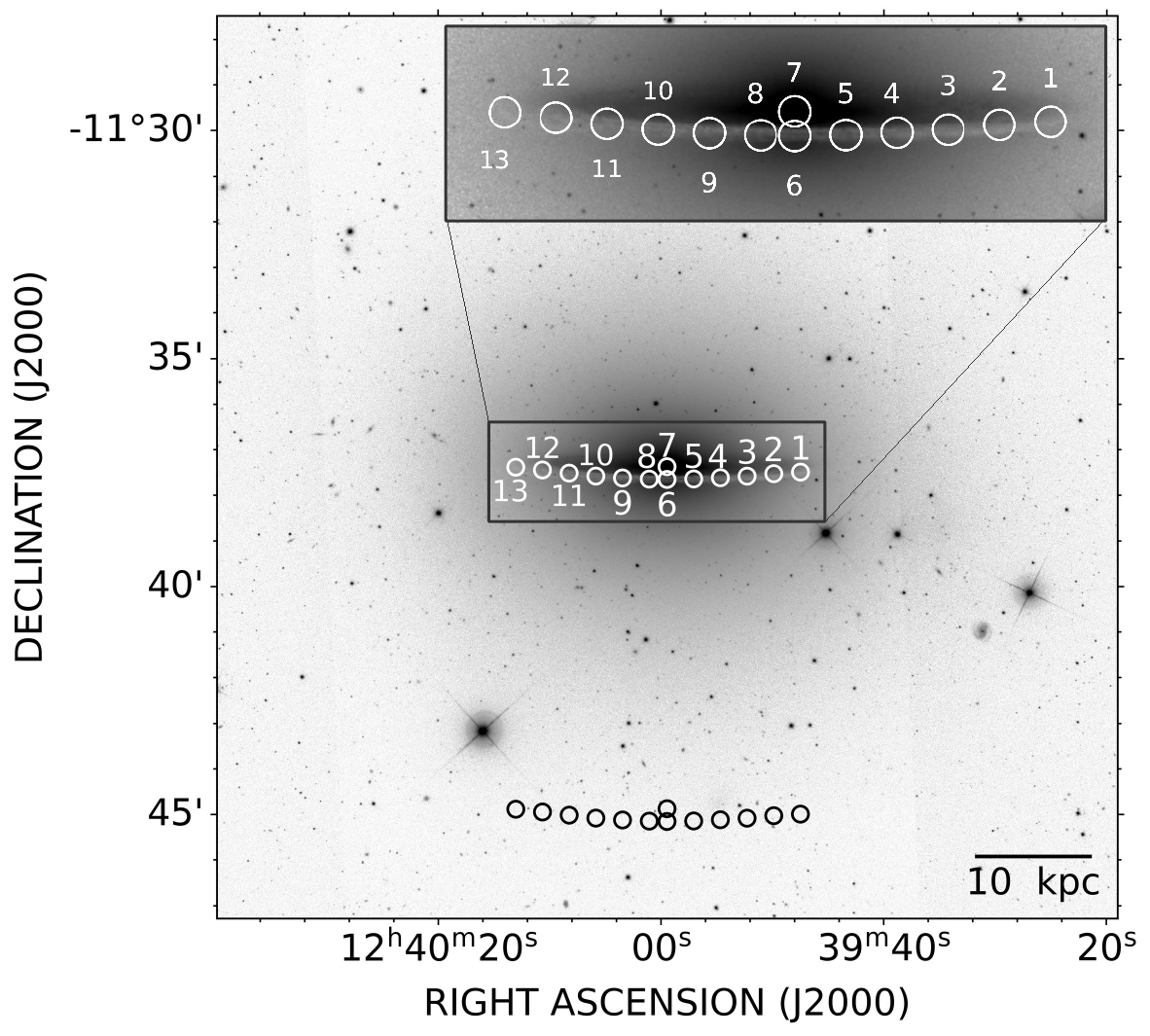,width=0.475\textwidth,angle=0}
\caption{SDSS $r$-band image of the $9.9^\prime\times9.9^\prime$ area centered at NGC~4594. The white circles are the location of the IRAM 30m beams with a diameter of $21.4^{\prime\prime}$ (HPBW at the CO~$J=1-0$ band). The center of the galaxy is at position ``7", while other observations are along the front side of the dust ring. The black circles to the south are the corresponding background positions for the PSW observations.
}\label{fig:IRAMbeam}
\end{center}
\end{figure}

We observed the central region of NGC~4594 with 12 pointings that are uniformly distributed along the front side of the dust ring (Fig.~\ref{fig:IRAMbeam}). We also added another pointing toward the nucleus of the galaxy. Each on-source exposure is $9.8\rm~minutes$, and all of the positions have at least two exposures. For some positions close to the edge of the dust ring, we at least doubled the exposure to achieve a higher signal-to-noise ratio (S/N). The observation log of each position is presented in Table~\ref{table:NGC4594obslog}. 

\subsection{Data reduction}\label{subsec:DataReduction}

We reduce the IRAM 30m data with the Continuum and Line Analysis Single-dish software (CLASS) package \footnote[1]{http://www.iram.fr/IRAMFR/GILDAS} version jan17a. The observations taken with the FTS backend often suffer from the ``platforming'' problem between individual units. After carefully masking the bad channels, we adopt the first-order polynomials to correct the platforms using the \textit{FtsPlatformingCorrection5.class} \footnote[2]{Provided by IRAM; \citet{Ute19}} script. We then smooth the spectra to a resolution of $\Delta v=10\rm~km~s^{-1}$ for $^{12}$CO $J=1-0$ line, which is typically sufficient to characterize the line shape. For the $^{13}$CO $J=1-0$ and $^{12}$CO $J=2-1$ lines, we resample them to a slightly higher resolution of $5\rm~km~s^{-1}$, in order to better resolve the weaker lines. We adopt a redshift of $z=0.003416$ from NED as the systemic redshift of NGC~4594. We also modify the main beam efficiency to the recommended values on the IRAM website: $B_{\rm eff}=78\%$ at 115~GHz and 59\% at 230~GHz. \footnote[3]{http://www.iram.es/IRAMES/mainWiki/Iram30mEfficiencies}

We use a polynomial model to fit the baseline, typically starting with a one-degree polynomial model while limiting the highest order to the third-degree in order to avoid overfittings. A large fraction of the scans were taken with a high water vapor and opacity (72\% with $\tau > 0.3$, 45\% with $\tau > 0.4$; Table~\ref{table:NGC4594obslog}), which could significantly affect the shape of the baseline in the spectra. We first filter the spectra with a RMS $>25\rm~mK$ at the E90 band and $>100\rm~mK$ at the E230 band. We further visually inspect the spectra and remove those with apparently strong standing waves. About 21\%, 36\%, and 69\% of the spectra for $^{12}$CO~$J=1-0$, $^{13}$CO~$J=1-0$, and $^{12}$CO~$J=2-1$ are removed because of the bad weather. 

We fit the spectra close to the emission line (typically within $\pm700\rm~km~s^{-1}$ from the systemic velocity of NGC~4594) with a one- or two-gaussian model plus a polynomial continuum. In the $^{12}$CO~$J=2-1$ spectra, we found some obvious ``dip'' at $200\rm~km~s^{-1}$ or $600\rm~km~s^{-1}$ in many spectra (Fig.~\ref{fig:12CO21spec}). We check the known absorption lines from the earth's atmosphere (using the \textit{Splatalogue} database \footnote[4]{https://splatalogue.online/advanced1.php}), compare the spectra extracted from different polarization components, weather conditions, and/or sky locations, but do not find any systematic biases caused by these issues. The origin of this ``dip'' is thus unknown, and likely artificial. In the following data analysis, we simply masked the velocity range of this feature in the $^{12}$CO~$J=2-1$ spectra.

\begin{figure*}
\begin{center}
\epsfig{figure=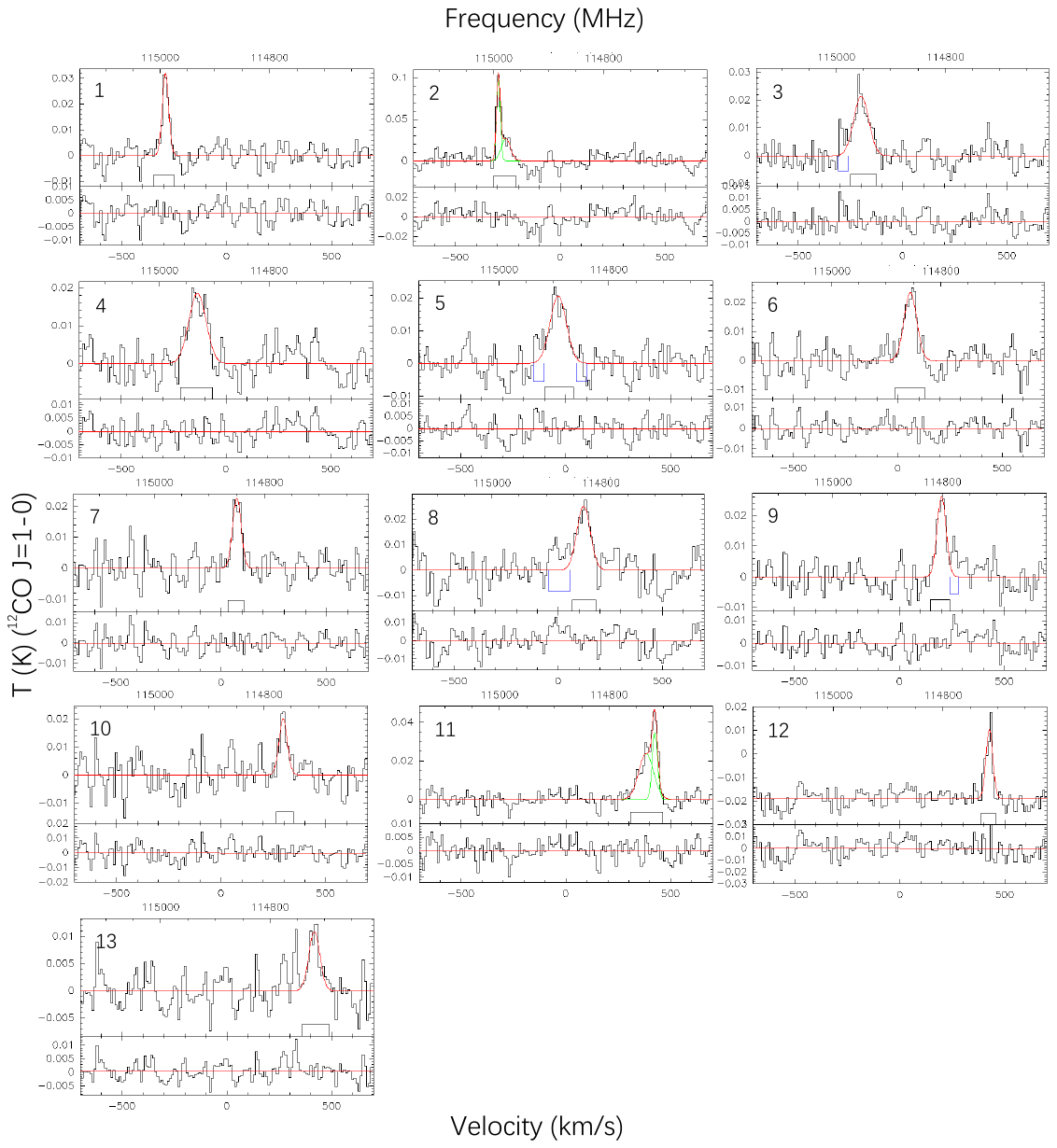, width=1.0\columnwidth,trim=20 180 0 50,clip}
\caption{Spectra around $^{12}$CO~$J=1-0$ at different locations as shown in Fig.~\ref{fig:IRAMbeam}. In the upper half of each panel, the $y$-axis is the main beam temperature after the correction of main beam forward efficiency. The lower half of each panel is the residual of the best-fit. The $x$-axis of all panels are the same ($114610-115147\rm~MHz$ for the top axis, or $\sim-700-+700\rm~km~s^{-1}$ for the bottom axis). The spectra have been binned to a velocity resolution of $10\rm~km~s^{-1}$, the zero velocity is set to the systematic velocity of the galaxy ($1024\rm~km~s^{-1}$). The green curve is the best-fit gaussian lines and the red line is the baseline. The black box shows the velocity range ($\Delta W$) used to fit the spectra. The blue boxes mask some low-significance features which could be artificial and may slightly affect the fitting. They are removed from the fitting and the calculation of the rms, while the exact definition of their frequency range will not significantly affect the results.
}\label{fig:12co10spec}
\end{center}
\end{figure*}

\begin{figure*}
\begin{center}
\epsfig{figure=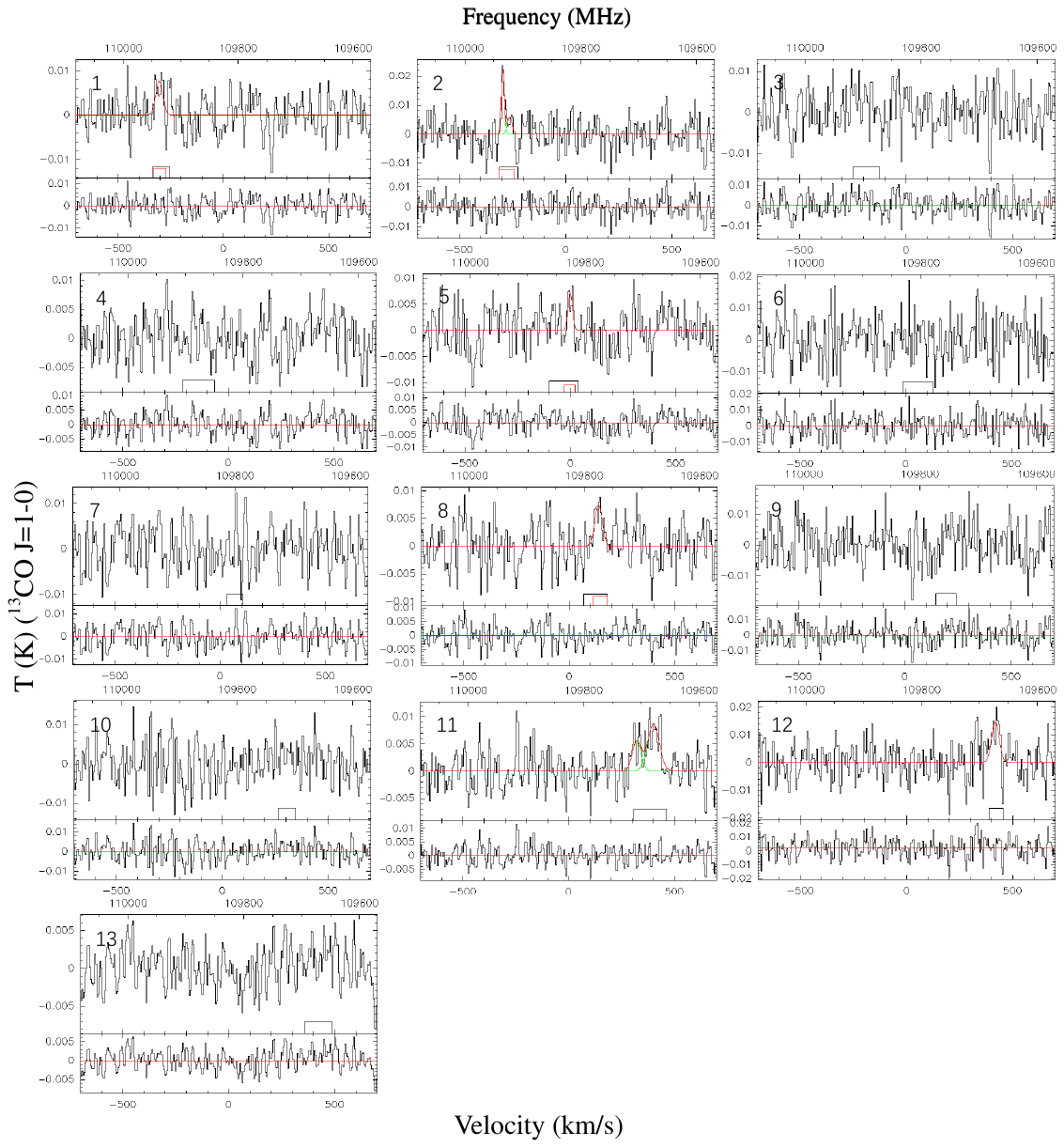, width=1.0\columnwidth,trim=20 180 0 50,clip}
\caption{Similar as Fig.~\ref{fig:12co10spec}, but for the $^{13}$CO~$J=1-0$ spectra. The $x$-axis range is $110082-1099569\rm~MHz$ and $\sim-700-+700\rm~km~s^{-1}$ for the top and bottom axes, respectively. The spectra have been binned to a velocity resolution of $5\rm~km~s^{-1}$. As $^{12}$CO~$J=1-0$ is often the strongest line and has been detected at all positions, we use the spectra of this line to choose the initial velocity range when fitting the spectra of the other two lines. This range can be slightly adjusted if necessary, and is plotted as boxes (black is the initial range, while red is the adjusted one).
}\label{fig:13CO10spec}
\end{center}
\end{figure*}

\begin{figure*}
\begin{center}
\epsfig{figure=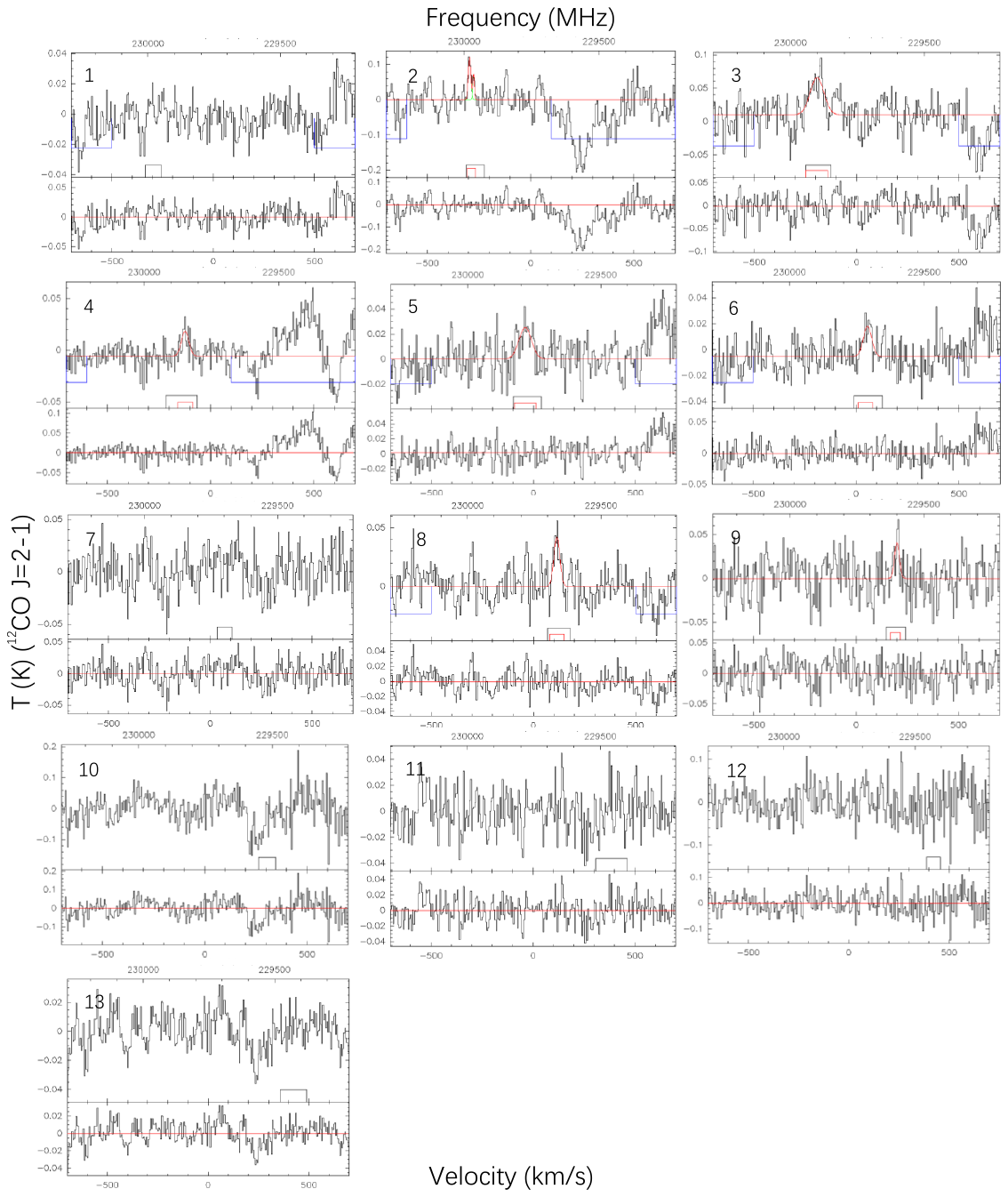, width=1.0\columnwidth,trim=20 130 0 50,clip}
\caption{Similar as Fig.~\ref{fig:12co10spec}, but for the $^{12}$CO~$J=2-1$ spectra. The $x$-axis range is $229216-230289\rm~MHz$ and $\sim-700-+700\rm~km~s^{-1}$ for the top and bottom axes, respectively. The spectra have been binned to a velocity resolution of $5\rm~km~s^{-1}$.}\label{fig:12CO21spec}
\end{center}
\end{figure*}

We calculate the intensity of different CO emission lines by integrating the gaussian profile over a velocity range of $\Delta W$, which is plotted in Figs.~\ref{fig:12co10spec}, \ref{fig:13CO10spec}, and \ref{fig:12CO21spec}. The uncertainty of integrated line intensity $\sigma$ is computed with a standard method as adopted in many previous works (e.g., \citealt{Braine93, Gao96}):
\begin{equation}\label{sigma}
 \sigma = T_{\rm rms}(\Delta W \Delta v)^{0.5}/(1-\Delta W/W)^{0.5},
\end{equation}
where $T_{\rm rms}$ is the rms around the line, $\Delta v$ is the velocity resolution, and $W$ is the entire velocity range to fit the baseline. In most of the cases, the $^{12}$CO~$J=1-0$ is the strongest line, and we have firmly detected it at all the observed positions. On the other hand, the $^{13}$CO~$J=1-0$ and $^{12}$CO~$J=2-1$ lines are not firmly detected at some positions. Whenever the lines are firmly detected, $\Delta W$ is determined based on the actual line shape. If however, the $^{13}$CO~$J=1-0$ and $^{12}$CO~$J=2-1$ lines are not firmly detected, we directly adopt the $\Delta W$ of $^{12}$CO~$J=1-0$. This is based on the assumption that the three lines have similar origin and velocity structure. If the integrated line intensity is less than three times of the integrated rms in the corresponding velocity range, we assume no detection of the line and calculate the 3$\sigma$ upper limit to the integrated line intensity instead. In case of the two-gaussian model, we use the intensity-weighted average velocity of the two gaussian components as the centroid velocity in the following analysis. 

When calculating the ratio between different emission lines, the beam dilution must be corrected at different frequencies. Following \citet{Li19}, we construct a 2-D ``image" of different CO lines based on the measurements at the isolated pointings, with the vertical extension equals to the beam size at the corresponding frequencies, while the horizontal extension of each pointing equals to the separation between the adjacent pointings. Such ``binned'' 2-D images of the intrinsic distribution of the line emissions are continuous along the disk, and could roughly account for the impact of the spilling out photons from the neighbouring beam (especially some beams covering bright point-like sources such as the AGN). It is certainly not an accurate description of the intrinsic distribution of the molecular gas, but since there seems to be no extremely bright point-like sources in NGC~4594 (as viewed from the IR images which often have a good correlation with the CO lines, e.g., Fig.~\ref{fig:VLAcontour}), we do not expect the specific assumptions on the intrinsic distribution of the molecular gas can significantly affect our results. We then use a gaussian kernel to convolve the higher frequency line image to the resolution of the lower frequency one, and use the convolved image to calculate the corresponding beam dilution factors. Here we use the gaussian kernel instead of a much more complicated stacked beam shape of different observations, which should not qualitatively change our results. The resultant beam dilution correction factors at different positions when calculating the $^{12}$CO~$J=1-0/^{13}$CO~$J=1-0$ and $^{12}$CO~$J=1-0/^{12}$CO~$J=2-1$ ratios are typically in the range of 0.95-0.99 and 0.47-0.55 (should be multiplied to the original line ratios), respectively. We summarize the measured or derived parameters of different CO emission lines at different positions in Table~\ref{table:COlinepara}.

\section{Results}\label{sec:results}

\subsection{Spatial distribution and total mass of molecular gas}\label{subsec:IntensityNH2}

\begin{figure}
\begin{center}
\epsfig{figure=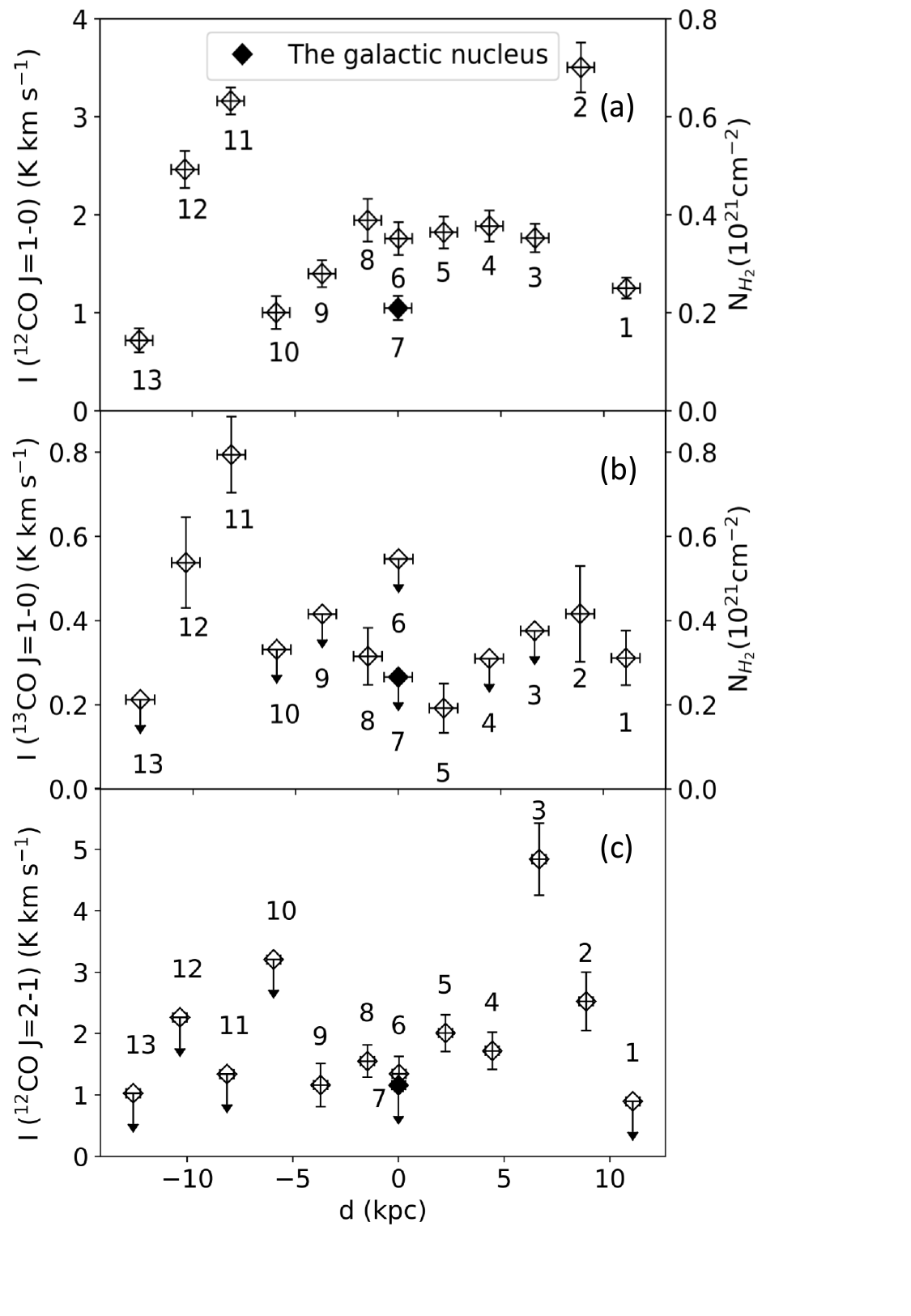, width=1.0\columnwidth,angle=0,trim=0 50 90 0,clip}
\caption{Integrated $^{12}$CO~$J=1-0$ (a), $^{13}$CO~$J=1-0$ (b), and $^{12}$CO~$J=2-1$ (c) line intensities along the front side of the dust ring of NGC~4594. The numbers besides the data points are the same as marked in Fig.~\ref{fig:IRAMbeam}. The $x$-axis is the projection distance to the minor axis. Position~``7" is the galactic nuclei so has the same distance to the minor axis as position~``6". The right axes of panels (a) and (b) are the molecular gas column density $N_{\rm H_2}$ calculated with the corresponding line. The vertical error bar is plotted at 1~$\sigma$ level while the upper limits are given at 3~$\sigma$ level. The horizontal error bar represents the beam size at the corresponding frequencies. 
}\label{fig:LineAreaProfile}
\end{center}
\end{figure}

\begin{figure}
\begin{center}
\epsfig{figure=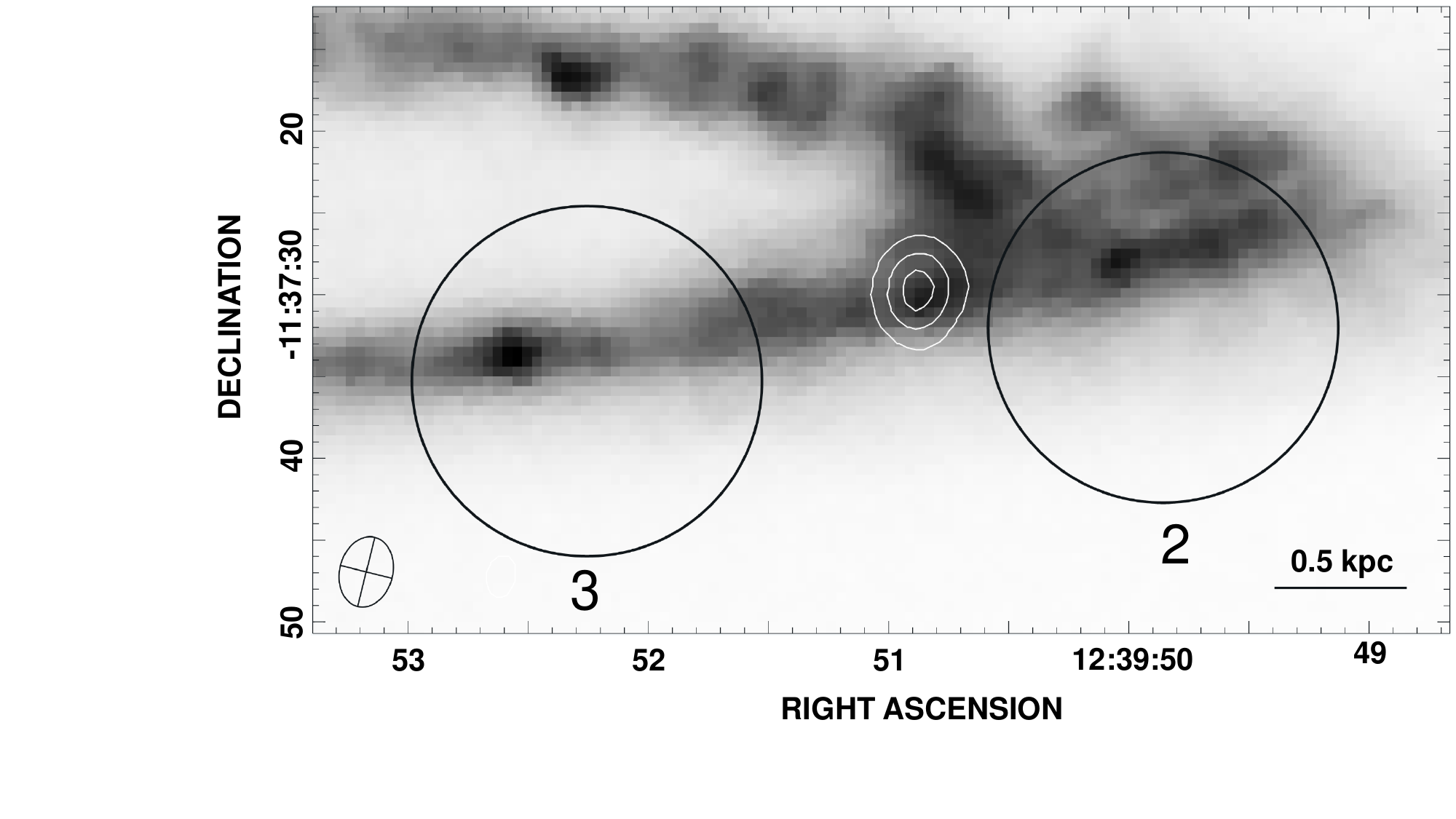, width=1.0\columnwidth,angle=0,trim=120 50 0 0,clip}
\caption{Spitzer 8~$\micron$ image of a $38^{\prime\prime}\times79^{\prime\prime}$ region on the western side of the dust ring of NGC~4594, covering positions ``2'' and ``3'' of our IRAM 30m observations (the two black circles are the same as the white circles in Fig.~\ref{fig:IRAMbeam}). The white contours are the radio source detected with the VLA B-configuration L-band (centered at 1.5~GHz) image (contour levels: 2.6$\sigma$, $7.1\sigma$ and $11.6\sigma$), with the beam shape denoted in the lower left corner.}
\label{fig:VLAcontour}
\end{center}
\end{figure}

We present in Fig.~\ref{fig:LineAreaProfile} the spatial distribution of the intensity of ${\rm ^{12}CO}$~$J=1-0$, ${\rm ^{13}CO}$~$J=1-0$, and ${\rm ^{12}CO}$~$J=2-1$ over a horizontal range of approximately $180^{\prime\prime}$ to $204^{\prime\prime}$ ($\approx11\rm~kpc$ to $-13\rm~kpc$) from the minor axis of the galaxy. The ${\rm ^{12}CO}$~$J=1-0$ line has been detected at all of the 12 positions along the front side of the dust ring and the nucleus of the galaxy (position~``7"), while the ${\rm ^{13}CO}$~$J=1-0$ and ${\rm ^{12}CO}$~$J=2-1$ lines are only firmly detected at 6/13 and 7/13 positions. The most significant signatures of these horizontal CO line intensity profiles are the two peaks in the ${\rm ^{12}CO}$~$J=1-0$ intensity profile at approximately the edge of the dust ring (at positions ``2", ``11", and ``12"). The enhanced CO intensity at these positions may be mainly caused by a projection effect. The locations of these two peaks are also roughly coincided with the horizontal profile of the \ion{H}{1} 21~cm line (two peaks at $\sim140^{\prime\prime}-150^{\prime\prime}$; \citealt{Bajaja84}), which apparently indicates that both the molecular and atomic gases have a ring-like structure. In particular, we also find significantly enhanced ${\rm ^{12}CO}$~$J=2-1$ emission at position ``3'' (Fig.~\ref{fig:LineAreaProfile}c). This unusually high ${\rm ^{12}CO}$~$J=2-1$ emission may be caused by the association with a patchy star forming region on the dust ring, as indicated in the IR images (the \emph{Spitzer} 8~$\rm \micron$ image as shown in Fig.~\ref{fig:VLAcontour}, as well as the \emph{Spitzer} $24\rm~\micron$ and \emph{Herschel} $70\rm~\micron$ images, all from \citealt{Sutter22}). Furthermore, we also discovered a bright radio source approximately between positions~``2" and ``3" with our latest VLA observations (Fig.~\ref{fig:VLAcontour}; will be presented in \citealt{Yang23}). This radio source is also apparently associated with an IR knot and possibly some H$\alpha$ knots as presented in \citet{Sutter22}. It could thus probably be a stellar radio source (e.g., pulsar or supernova remnant) from a compact SF region, although we cannot rule out the possibility that it is a background source. Some of these SF regions may occasionally fall into the main or side lobes of our IRAM 30m observations, which may also contribute to the enhanced ${\rm ^{12}CO}$~$J=2-1$ emission.

We further derive the molecular gas column density $N_{\rm H_2}$ using the ${\rm ^{12}CO}$ and ${\rm ^{13}CO}$~$J=1-0$ line intensities, assuming a filling factor of unity within the main lobe. The corresponding conversion factor is defined as $X\equiv N_{\rm H_2}/I_{\rm CO}$, where $I_{\rm CO}$ is the integrated CO line intensity in units of $\rm K~km~s^{-1}$, while $N_{\rm H_2}$ is the molecular hydrogen column density in units of $10^{21}\rm~cm^{-2}$. $N_{\rm H_2}$ depends on the filling factor $f$ in the manner of $N_{\rm H_2}\propto f^{-1}$. The $X$ factor is affected by the physical and chemical properties of the molecular clouds (e.g., temperature, velocity dispersion, metallicity, density, etc.), but is typically in a relatively narrow range (e.g., \citealt{Shetty11b}). In this paper, we simply adopt a constant $X$ factor of $0.2\times10^{21}\rm~cm^{-2}/(K~km~s^{-1})$ for ${\rm ^{12}CO}$~$J=1-0$ (e.g., \citealt{Shetty11a,Bolatto13}). Since ${\rm ^{12}CO}$~$J=1-0$ is often optically thick at high $H_2$ column densities, we also estimate $N_{\rm H_2}$ based on the weaker ${\rm ^{13}CO}$~$J=1-0$ line which in most of the cases is optically thin. We adopt an $X$ factor of $1.0\times10^{21}\rm~cm^{-2}/(K~km~s^{-1})$ for ${\rm ^{13}CO}$~$J=1-0$, which has a typical uncertainty of a factor of two (e.g., \citealt{Cormier2018}). This ${\rm ^{13}CO}$~$J=1-0$ $X$ factor is based on a calibration of the molecular gas mass from IR observations of the dust emission (assuming a dust-to-molecular gas ratio). It has been suggested that ${\rm ^{13}CO}$~$J=1-0$ is a more reliable tracer than ${\rm ^{12}CO}$~$J=1-0$ in regions with a higher fraction of dense gas, but in general a less reliable predictor of the bulk molecular gas in normal galaxies due to the presence of a large diffuse gas phase (e.g., \citealt{Cormier2018}). We caution that NGC~4594 may have a lower $X$ factor due to its possibly high metallicity, which may bring in additional uncertainties to the calculation.

When calculating the total molecular gas mass of the galaxy using the ${\rm CO}~J=1-0$ lines, we assume that most of the molecular gas of NGC~4594 is distributed in the dust ring (as revealed in \citealt{Sutter22}) and the vertical extension of the molecular gas ring equals to the main beam size at the frequency of the ${\rm CO}~J=1-0$ lines ($\approx21.4^{\prime\prime}$). For the molecular gas mass calculated at the positions where the ${\rm ^{13}CO}$~$J=1-0$ line is not detected, we use the measured upper limits to characterize its probability distribution (assuming gaussian distribution of a $\geq0$ mass characterized with a best-fit mass of zero and the measured upper limit). We adopt $5\times10^4$ boot-strap samples to determine the distributions of each measurement data and calculate the median value and uncertainty of the total mass from different positions. Considering that the disk is not completely edge-on and the back half of the ring is not covered by our observations, we simply assume the total mass contained in the molecular ring is about twice of our directly measured value. Adding the mass measured at the nuclear region (position ``7"), we obtain the total molecular gas measured from the ${\rm ^{12}CO}~J=1-0$ line is $M_{\rm H_2} = (4.0\pm0.1)\times10^{8}\rm~M_\odot$, while that from the ${\rm ^{13}CO}~J=1-0$ line is $M_{\rm H_2} = 2.9^{+0.3}_{-0.2}\times10^{8}\rm~M_\odot$.

The above estimate of $N_{\rm H_2}$ is directly linked to the assumption on the spatial distribution of the gas, which is characterized with the filling factor $f$. However, the total mass is calculated based on the average column density within the beam, so not affected by $f$. According to the spatially resolved ALMA ${\rm ^{12}CO}$~$J=1-0$ image, the molecular gas is mostly distributed in the dust ring, although the fine structures are different from the ionized gas tracers such as the H$\alpha$ and [\ion{C}{2}] $158\rm~\micron$ lines \citep{Sutter22}. This is also supported by the IR images which show the prominent dust ring (Fig.~\ref{fig:VLAcontour}). Therefore, the above assumption of the geometry of the spatial distribution of the molecular gas is in general correct. However, since our IRAM 30m beam sizes at the frequencies of the CO~$J=1-0$ lines are larger than the thickness of the molecular gas ring (comparable to the dust ring), we expect the actual filling factor is $f<1$ and the $N_{\rm H_2}$ estimated here is a lower limit. Nevertheless, for a uniform comparison to other CO-CHANGES galaxies and also because we do not have a reliable way to quantitatively calculate $f$, we do not adopt any additional corrections related to the filling factor.

\subsection{CO line ratios and physical conditions of molecular gas}\label{subsec:RatioPara}

\begin{figure}
\begin{center}
\epsfig{figure=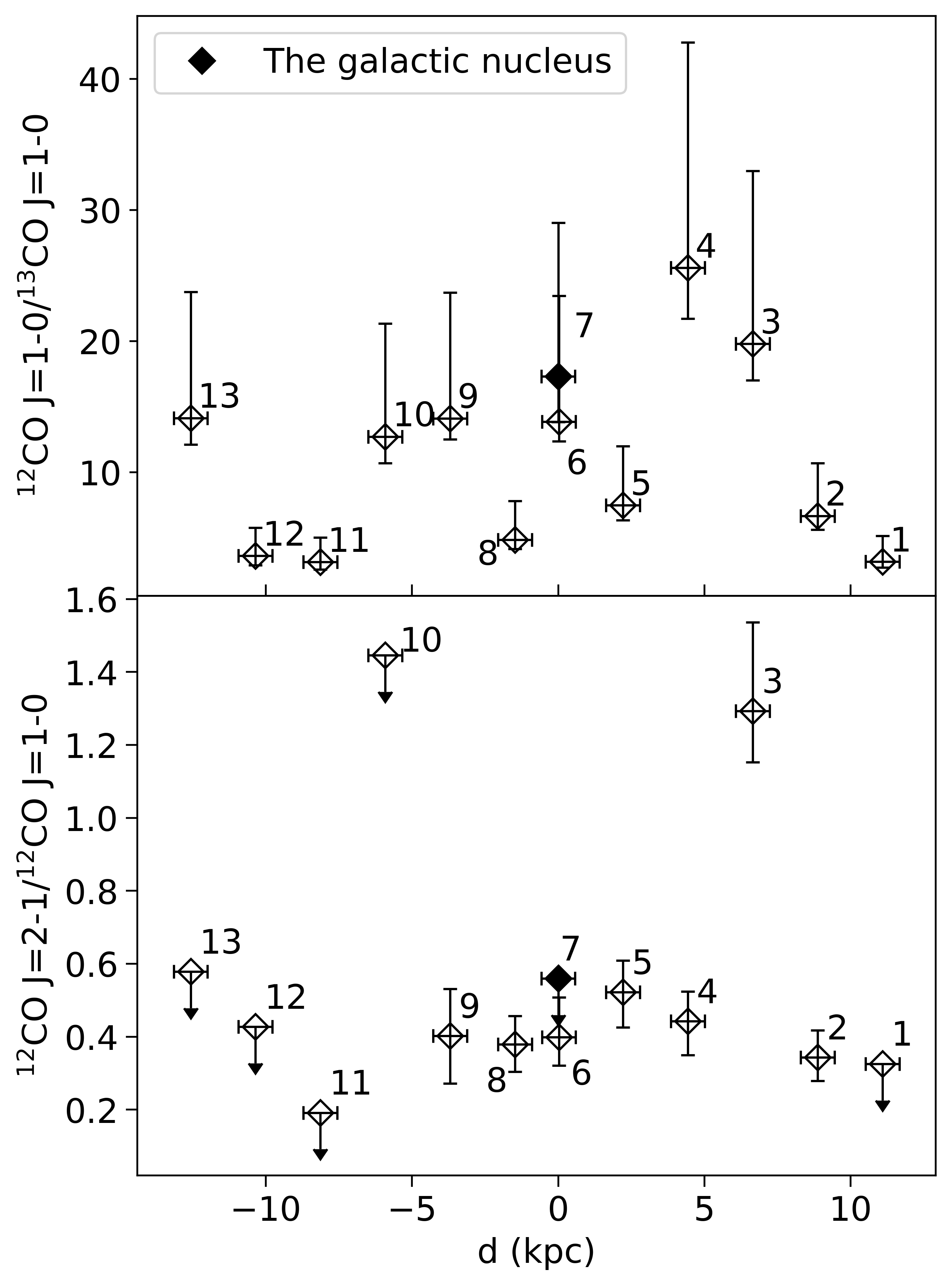, width=1.0\columnwidth}
\caption{Horizontal distributions of the CO line ratios: (a) $^{12}$CO/$^{13}$CO~$J=1-0$ line ratio; (b) $^{12}$CO~$J=2-1$/$J=1-0$. Both line ratios have been corrected for beam dilution.}
\label{fig:RatioProfile}
\end{center}
\end{figure}

The physical properties of the molecular gas can be characterized with the CO line ratios. We present in Fig.~\ref{fig:RatioProfile} the spatial distributions of the $^{12}$CO/$^{13}$CO~$J=1-0$ and $^{12}$CO~$J=2-1$/$J=1-0$ intensity ratios after correcting for the beam dilution (\S\ref{subsec:DataReduction}). For the positions with undetected $^{13}$CO~$J=1-0$ or $^{12}$CO~$J=2-1$ lines ($^{12}$CO~$J=1-0$ has been detected at all positions), we use the same method adopted in calculating the molecular gas mass in \S\ref{subsec:IntensityNH2} to calculate the related parameters. The median values of the $^{12}$CO/$^{13}$CO~$J=1-0$ and $^{12}$CO~$J=2-1$/$J=1-0$ line ratios on the dust ring (excluding the nuclear region) are $9.2^{+3.4}_{-2.1}$ and $0.34^{+0.05}_{-0.04}$, respectively. These line ratios are consistent with those measured from quiescent star forming galaxies (e.g., \citealt{Tan11,Alatalo15a,Cormier2018}). The $^{12}$CO/$^{13}$CO~$J=1-0$ and $^{12}$CO~$J=2-1$/$J=1-0$ line ratios of the nuclear region are $17.3^{+11.7}_{-3.5}$ and $\textless~0.56$, respectively. Most of the measured CO line ratios on the dust ring are consistent with each other within the uncertainties, except for ``Position 3'' which has a significantly larger $^{12}$CO~$J=2-1$/$J=1-0$. As discussed in \S\ref{subsec:IntensityNH2} and shown in Fig.~\ref{fig:LineAreaProfile}c, this may be caused by the strong $^{12}$CO~$J=2-1$ emission related to some disk SF regions falling in the beam.

\begin{figure}
\begin{center}
\epsfig{figure=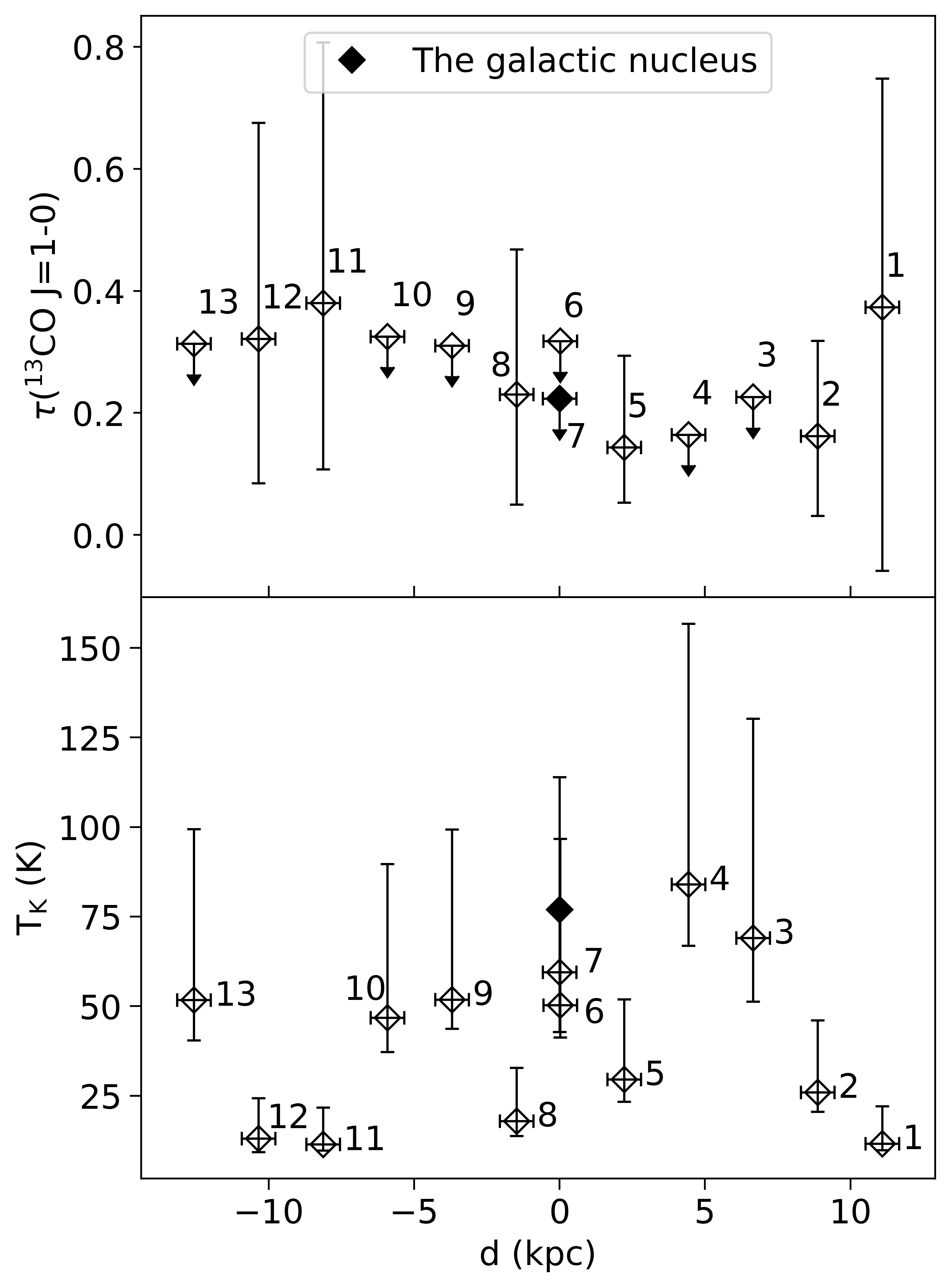, width=1.0\columnwidth}
\caption{Horizontal distributions of (a) Optical depth of $^{13}$CO ($\tau_{\rm ^{13}CO}$); (b) Kinetic temperature ($T_{\rm K}$) of the molecular gas under the LTE assumption.}\label{fig:ParaProfile}
\end{center}
\end{figure}

We further calculate the optical depth $\tau$ and kinetic temperature $T_{\rm K}$ of the molecular gas using the $^{12}$CO/$^{13}$CO~$J=1-0$ line ratio, based on several assumptions: (1) $^{12}$CO~$J=1-0$ and $^{13}$CO~$J=1-0$ have the same excitation temperature and filling factor; (2) the molecular gas is under LTE (the local thermal equilibrium) condition; (3) $\tau_{\rm ^{12}CO} \gg$ 1 and $\tau_{\rm ^{13}CO} \ll$ 1. These assumptions are in general correct in a galaxy like NGC~4594 with no extreme conditions (i.e., very intense nuclear starburst; e.g., \citealt{Tan11,Li19}). Based on these assumptions, we can calculate the optical depth of $^{13}$CO~$J=1-0$ using the following equation:
\begin{equation}\label{equi:tau13CO}
\tau(\rm ^{13}CO)=-\ln[1-\frac{I_{\rm^{13}CO~(J=1-0)}}{I_{\rm ^{12}CO~(J=1-0)}}].
\end{equation}
On the other hand, $T_{\rm K}$ could be calculated with the following equation:
\begin{equation}\label{equi:NH213CO}
\frac{N_{\rm H_2}}{\rm cm^{-2}}=2.25\times10^{20}[\frac{\tau_{\rm ^{13}CO}}{1-e^{-\tau_{\rm ^{13}CO}}}]\frac{I_{\rm ^{13}CO~J=1-0}}{1-e^{-5.29/T_{\rm ex}}},
\end{equation}
assuming that the $^{13}$CO abundance ${\rm [^{13}CO]/[H_2]}$ is $8\times10^{-5}/60$ \citep{Frerking82}. The factor ``60'' is the $\rm ^{12}CO$/$\rm ^{13}CO$ abundance ratio, which could distribute in a wide range (e.g., \citealt{Liszt17}). Here we adopt a value of 60 following \citet{Cormier2018}. It is possible that the $^{13}$CO abundance varies at different positions, but this should have a limited impact on our final results, given that the SF and molecular gas properties do not change dramatically across the disk of NGC~4594. Under LTE, the excitation temperature $T_{\rm ex}$ equals to the kinetic temperature $T_{\rm K}$. $\tau(\rm ^{13}CO)$ and $T_{\rm K}$ measured at different positions are listed in Table~\ref{table:COlinepara}, while their spatial distributions are shown in Fig.~\ref{fig:ParaProfile}. Compared to similar measurements in other galaxies (e.g., \citealt{Tan11}), the typical values of $\tau(\rm ^{13}CO)\sim0.1-0.5$ and $T_{\rm K}\sim10-80\rm~K$ do not indicate a significant heating by external sources such as SF [typically has smaller $\tau(\rm ^{13}CO)$ and higher $T_{\rm K}$]. This is consistent with the above assumption that there is no extreme conditions of the molecular gas in NGC~4594.

\subsection{Position velocity diagram}\label{subsec:PVD}

\begin{figure}
\begin{center}
\epsfig{figure=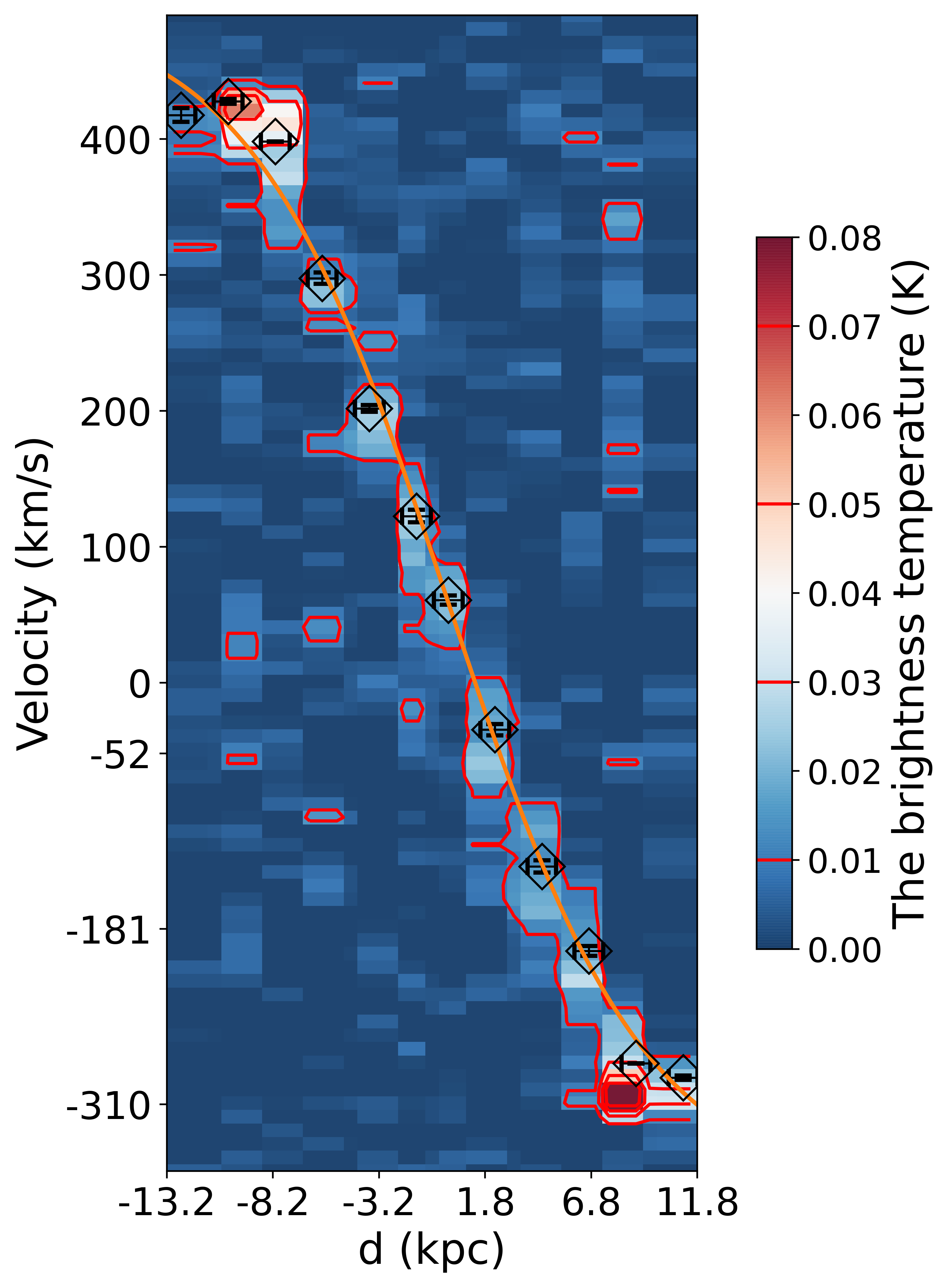, width=1.0\columnwidth}
\caption{The position-velocity diagram of $^{12}$CO~$J=1-0$ of NGC~4594. Most of the pixel size in the horizontal direction equals to half of the beam size, except for the central three which equals to the separation between them. The velocity resolution is $10\rm~km~s^{-1}$. The color bar shows the brightness temperature and the contour levels are 0.01, 0.03, 0.05, and 0.07 $\rm K$. The black diamond locates at the centroid velocity of each position. The orange curve is the best-fit rotation curve with Eq. (\ref{equi:RC}). 
}\label{fig:PVdiagram}
\end{center}
\end{figure}

We show the $^{12}$CO~$J=1-0$ position-velocity diagram (PVD) along the front side of the dust ring of NGC~4594 in Fig.~\ref{fig:PVdiagram}. The eastern side (with positive velocity and negative distance) is the following side, while the opposite side is the preceding side \citep{Bajaja84}. There is a clear trend of a flattened rotation curve at the outermost positions, although the flattened part is not well characterized with our data. Detailed analysis of the CO rotation curve of the CHANG-ES sample, as well as comparisons with the rotation curves of multi-phase gases, will be presented in follow-up papers. We herein only fit the rotation curve with a hyperbolic tangent function plus a linear function in order to roughly estimate the peak velocity of the rotation curve $V_{\rm max}$ \citep{Yoon21}:
\begin{equation}\label{equi:RC}
V(r)=V_{\rm max}\rm tanh(\frac{r}{R_{\rm t}})+s_{\rm out}r,
\end{equation}
where $s_{\rm out}$ is the slope of the rotation curve when $r\gg R_{\rm t}$, $R_{\rm t}$ is the turnover radius where the hyperbolic tangent term begins to flatten. When $s_{\rm out}=0$, $V_{\rm max}$ is the maximum velocity of the rotation curve. Since $s_{\rm out}$ is often small and the hyperbolic tangent component often dominates close to the galactic center, we adopt $V_{\rm max}$ as the maximum rotation velocity, which has a best-fit value of $374\pm 32\rm~km~s^{-1}$. Adopting the inclination angle of $i\approx79^\circ$ (Table~\ref{table:NGC4594}), the inclination corrected maximum rotation velocity will be $V_{\rm max,corr}=381\pm 32\rm~km~s^{-1}$. Such a large maximum rotation velocity indicates that the dark matter halo mass is $M_{\rm 200}\gtrsim10^{13}\rm~M_\odot$ (e.g., \citealt{Salucci07}; $M_{\rm 200}$ is the dark matter halo mass enclosed in a radius where the average density is 200 times the critical density of the universe).

\section{Discussions}\label{sec:discussion}

As introduced in \S\ref{sec:Introduction}, NGC~4594 is the most massive spiral galaxy in the local universe and locates in an isolated environment. It thus provides us with a unique opportunity to witness the growth of such a massive galaxy in relative isolation. In this section, we compare our measured molecular gas properties of NGC~4594 to other galaxy samples observed with similar instruments, aiming at understanding the role of molecular gas in the gas circulation between such a massive isolated galaxy and its environment.

\subsection{Galaxy samples used for comparison}\label{subsec:galaxysamples}

The galaxy samples included for comparison in the present paper are selected from the ``CO Multi-line Imaging of Nearby Galaxies (COMING)'' sample \citep{Muraoka16} and the ATLAS$\rm^{3D}$ sample \citep{Young11}. Most of the galaxies from the COMING sample are late-type, while the ATLAS$\rm^{3D}$ sample is mostly comprised of early-type galaxies. The COMING project is comprised of the Nobeyama 45m and IRAM 30m observations of the $^{12}$CO and $^{13}$CO $J=1-0$, and $^{12}$CO $J=2-1$ lines of 147 galaxies with high far-IR intensities \citep{Muraoka16}. \citet{Yajima21} provided spatially resolved measurements of the CO lines of 16 galaxies from the COMING sample with all of these three lines detected. We calculate two sets of each of the CO line ratios of each galaxy in the COMING sample: values directly measured from the nuclear region and the median values measured from non-nuclear regions. The parameters from these two types of regions could be compared to those of the corresponding regions of NGC~4594.

The ATLAS$\rm^{3D}$ project studies the $^{12}$CO $J=1-0$ and $J=2-1$ lines in 260 early-type galaxies with the IRAM 30m telescope, but only 41 galaxies have both lines firmly detected \citep{Young11}. We also obtain the $^{13}$CO $J=1-0$ measurements of 18 ATLAS$\rm^{3D}$ galaxies from some later IRAM 30m observations \citep{Crocker12}. However, \citet{Crocker12} only measured the CO line properties from the nuclear region of these 18 galaxies. These measurements will therefore only be compared to the corresponding measurements of the nuclear region of NGC~4594. It is noteworthy that for the ATLAS$\rm^{3D}$ sample, \citet{Young11} adopted a different $X$-factor of $X=3.0\times10^{20}\rm~cm^{-2}/(K~km~s^{-1})$ for $^{12}$CO~$J=1-0$ than that adopted in the present paper. Therefore, we recalculate their results using the same $X$ factor as the present paper for a uniform comparison.

We also compare to another massive spiral galaxy, NGC~5908, which is similar as NGC~4594 in stellar mass and morphological type, but more than one order of magnitude higher in molecular or total cold gas content \citep{Li19}. Parameters of NGC~5908 are directly quoted from \citet{Li19}.

The SF properties of the ATLAS$\rm^{3D}$ sample are obtained from \citet{Davis14}, while those of the COMING sample are from \citet{Yajima21}. Both works use the WISE $22~\micron$ map and GALEX FUV map to calibrate the SFR, following the method described in \citet{Calzetti07}. For NGC~4594, we calculate the SF properties using the SFR estimated from both the H$\alpha$ and WISE $22~\micron$ fluxes \citep{Vargas19}, as well as the $22~\micron$ SF radius from \citet{Wiegert15}.
Similar as in \citet{Li13a}, we use the morphological type code ($TC$) obtained from the HyperLeda database (\url{http://leda.univ-lyon1.fr/}) to quantitatively classify the morphology of the galaxies. We define galaxies with $TC\geq1.5$ as late-type, while those with $TC<1.5$ as early-type.

\begin{figure}
\begin{center}
\epsfig{figure=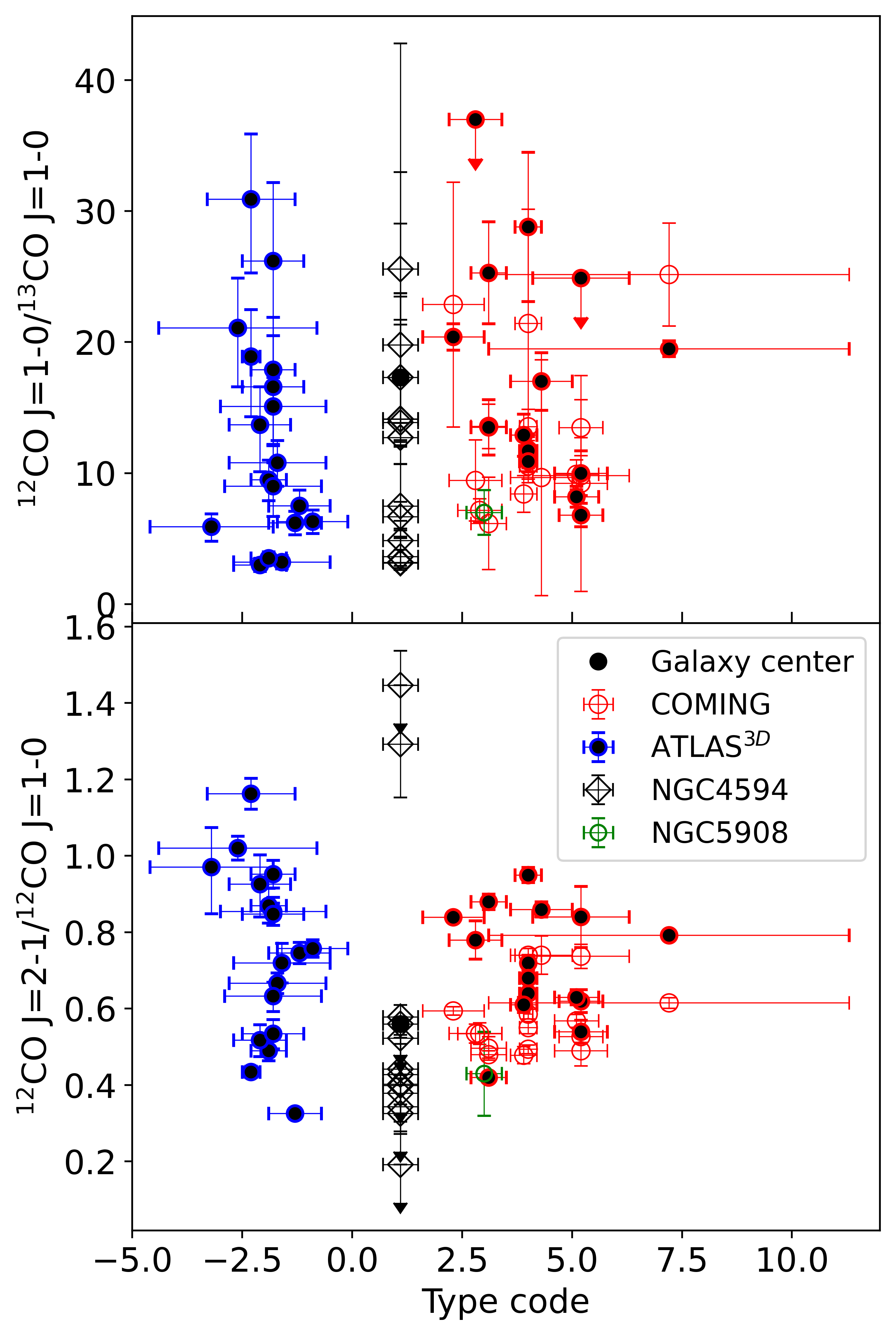, width=1.0\columnwidth}
\caption{The morphological type code and $^{12}$CO/$^{13}$CO~$J=1-0$ and $^{12}$CO~$J=2-1$/$J=1-0$ line ratios of the galaxies included for comparison (as denoted at the upper right corner of the lower panel). The type code of the galaxies are obtained from HyperLeda, while the line ratios are obtained from different references as described in \S\ref{subsec:galaxysamples}. Open symbols are calculated in non-nuclear regions, while the filled symbols are calculated only in the nuclear regions. Galaxies obtained from the ATLAS$\rm^{3D}$ sample has only one observation per galaxy toward the center \citep{Crocker12}, so are regarded as measurements from the nuclear region.
}\label{fig:lineratioandtypecode}
\end{center}
\end{figure}

As shown in Fig.~\ref{fig:lineratioandtypecode}, NGC~4594 has a $TC$ between early- and late-type galaxies from the ATLAS$\rm^{3D}$ and COMING samples, respectively. The $^{12}$CO~$J=2-1$/$J=1-0$ line ratio of NGC~4594 is among the lowest (except for two positions) of all the galaxies included for comparison. However, we caution that the selected galaxies from the COMING and ATLAS$\rm^{3D}$ samples may be biased to the most gas-rich ones in the corresponding morphological types, as we need all the three CO lines to be detected to calculate the line ratios. This bias may affect the comparison below. On the other hand, the difference between the CO line ratios in the nuclear and outer regions appears more significant, with the galactic nuclear regions always have higher $^{12}$CO~$J=2-1$/$J=1-0$ but lower $^{12}$CO/$^{13}$CO~$J=1-0$ than the outer regions, indicating clearly different physical conditions of the molecular gas in these two types of regions of a galaxy.

\subsection{CO line ratios}\label{subsec:DiscussCOLineRatio}

The ratios of different CO emission lines are quite informative of the physical conditions of the molecular gas. In this section, we compare the $^{12}$CO/$^{13}$CO~$J=1-0$ and $^{12}$CO~$J=2-1$/$J=1-0$ line ratios measured at different positions of NGC~4594 to the selected galaxies from the COMING and ATLAS$\rm^{3D}$ samples, as well as a gas richer massive spiral galaxy NGC~5908 (\S\ref{subsec:galaxysamples}).

The $^{12}$CO/$^{13}$CO~$J=1-0$ line ratio is affected by the optical depth and the abundance of $^{13}$CO molecules (e.g. \citealt{Alatalo15a}). The $^{13}$CO abundance elevates mainly through isotope exchange reactions in cold interstellar clouds and the CN cycle in intermediate-mass stars (\citealt{Sage91}). The combined effect is a suppress of $^{13}$CO or a higher $^{12}$CO/$^{13}$CO~$J=1-0$ line ratio in warm molecular clouds and active star formation regions (e.g., \citealt{Jim17}). On the other hand, the optical depth could be affected by many factors, such as the variation in temperature and density of the molecular gas, or the presence of a diffuse, non-self-gravitating gas phase. The $^{12}$CO emission mainly arises from the warm diffuse inter-clumpy regions, while the $^{13}$CO emission mainly arises from the dense molecular cores. When the molecular clouds are heated by various mechanisms in active star forming regions, the optical depth of the $^{12}$CO-rich envelope significantly decreases, allowing more $^{12}$CO photons to escape. The result is similar as the abundance effect, i.e., a higher $^{12}$CO/$^{13}$CO~$J=1-0$ line ratio in a more intensely star forming environment (e.g., \citealt{Tan11, Jim17, Cormier2018}). 

The $^{12}$CO~$J=2-1$/$J=1-0$ line ratio is basically influenced by the molecular gas temperature and optical depth, thus the structure and heating sources of the molecular clouds. The molecular clouds in the Milky Way could be divided into three categories based on their $^{12}$CO~$J=2-1$/$J=1-0$ line ratio: the low ratio gas with the $^{12}$CO~$J=2-1$/$J=1-0$ line ratio $\textless$~0.7, the high ratio gas with the $^{12}$CO~$J=2-1$/$J=1-0$ line ratio $=0.7-1.0$, and the very high ratio gas with the $^{12}$CO~$J=2-1$/$J=1-0$ line ratio $\textgreater$~1 \citep{Hasegawa97}. The low ratio gas comes from the extended low-density envelope of molecular clumps. The high ratio gas originates from the highly confined molecular clumps with a steep density gradient and a thin CO emitting envelope (\citealt{Penaloza17}). Their classification is mostly affected by the optical depth effect. The very high ratio gas is often significantly heated by external sources, such as UV photons from young stars or shocks from SNe and stellar winds, which could significantly increase the temperature and high-$J$ CO emission of the cloud (e.g., \citealt{Bolatto13}).

\begin{figure}
\begin{center}
\epsfig{figure=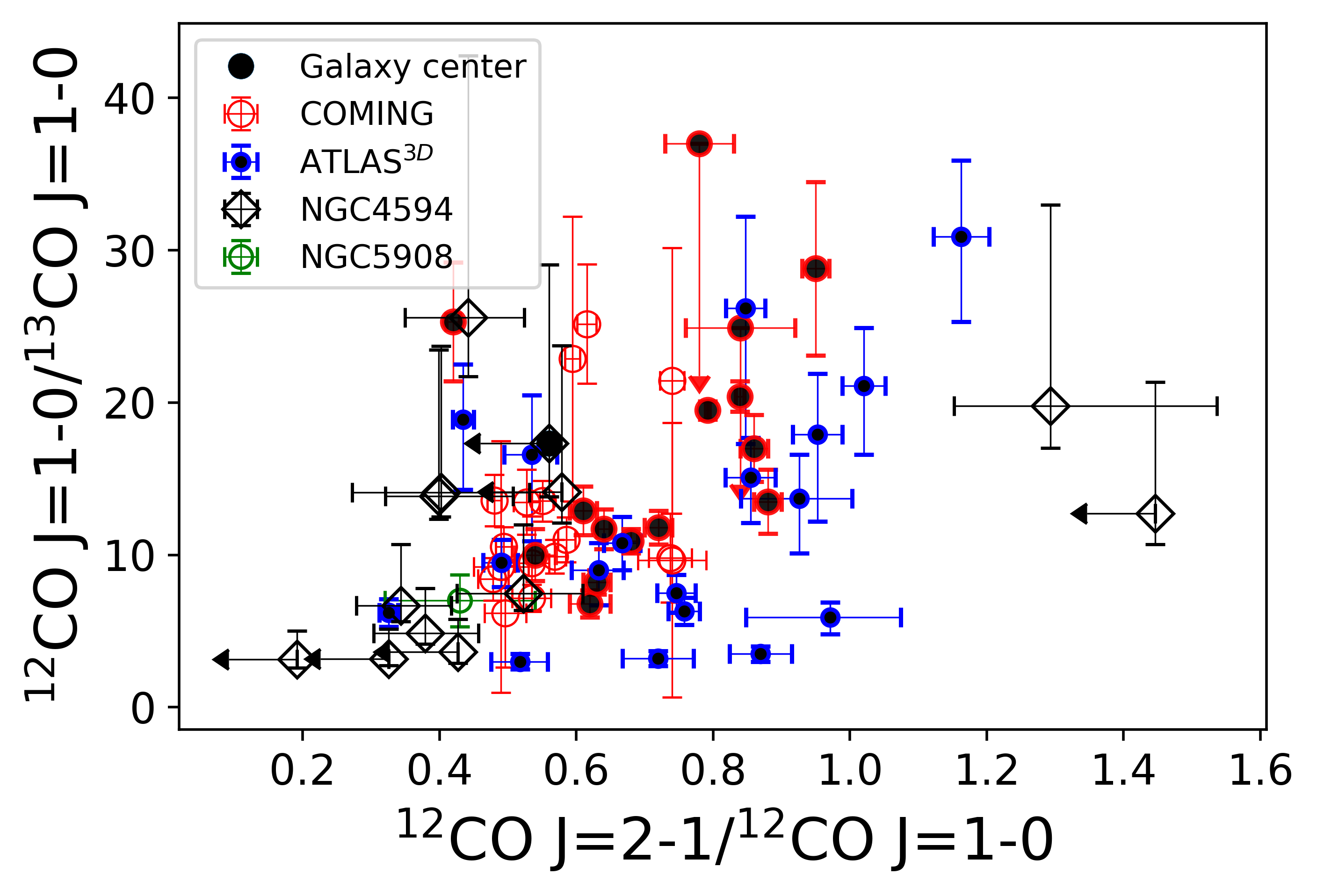, width=1.0\columnwidth}
\caption{The ``color-color diagram'' of different CO line ratios. Symbols of different galaxy samples are the same as in Fig.~\ref{fig:lineratioandtypecode}.
}\label{fig:ratio&ratio}
\end{center}
\end{figure}

Fig.~\ref{fig:ratio&ratio} shows the comparison of the CO line ratios of NGC~4594 to the COMING and ATLAS$\rm^{3D}$ samples on the CO line ``color-color diagram''. As described in \S\ref{subsec:galaxysamples}, the ATLAS$\rm^{3D}$ sample only has single-pointing observations of each galaxy which are adopted as parameters of the nuclear region, while the COMING sample includes spatially resolved observations covering both the nuclear and outer regions. These observations of different regions should be compared to our observations of the corresponding regions in NGC~4594. There is in general a positive correlation between the $^{12}$CO~$J=2-1$/$J=1-0$ and the $^{12}$CO/$^{13}$CO~$J=1-0$ line ratios, which is most likely a result of the star formation effect. As discussed above, a combined effect of a more active star formation will result in both a higher $^{12}$CO~$J=2-1$/$J=1-0$ and $^{12}$CO/$^{13}$CO~$J=1-0$. As we only select galaxies with reliable observations of all of the three CO lines from the COMING and ATLAS$\rm^{3D}$ samples, the selected subsamples may be biased to the most gas-rich and SF active ones (\S\ref{subsec:galaxysamples}). This makes NGC~4594 locate at the lower left corner of Fig.~\ref{fig:ratio&ratio}. The measured line ratios of NGC~4594 clearly stay in the area of quiescent galaxies (e.g., \citealt{Tan11,Alatalo15a,Cormier2018}), which is consistent with its low SFR surface density ($\Sigma_{\rm SFR}$ in Table~\ref{table:NGC4594}). These line ratios are even lower than those measured for NGC~5908, another very massive spiral galaxy with a huge stellar budge but higher molecular gas content \citep{Li19}, indicating that NGC~4594 is extremely gas poor and inactive in SF, although a recent study reveals some AGN activity and associated large-scale radio jet relics \citep{Yang23}.

\begin{figure}
\begin{center}
\epsfig{figure=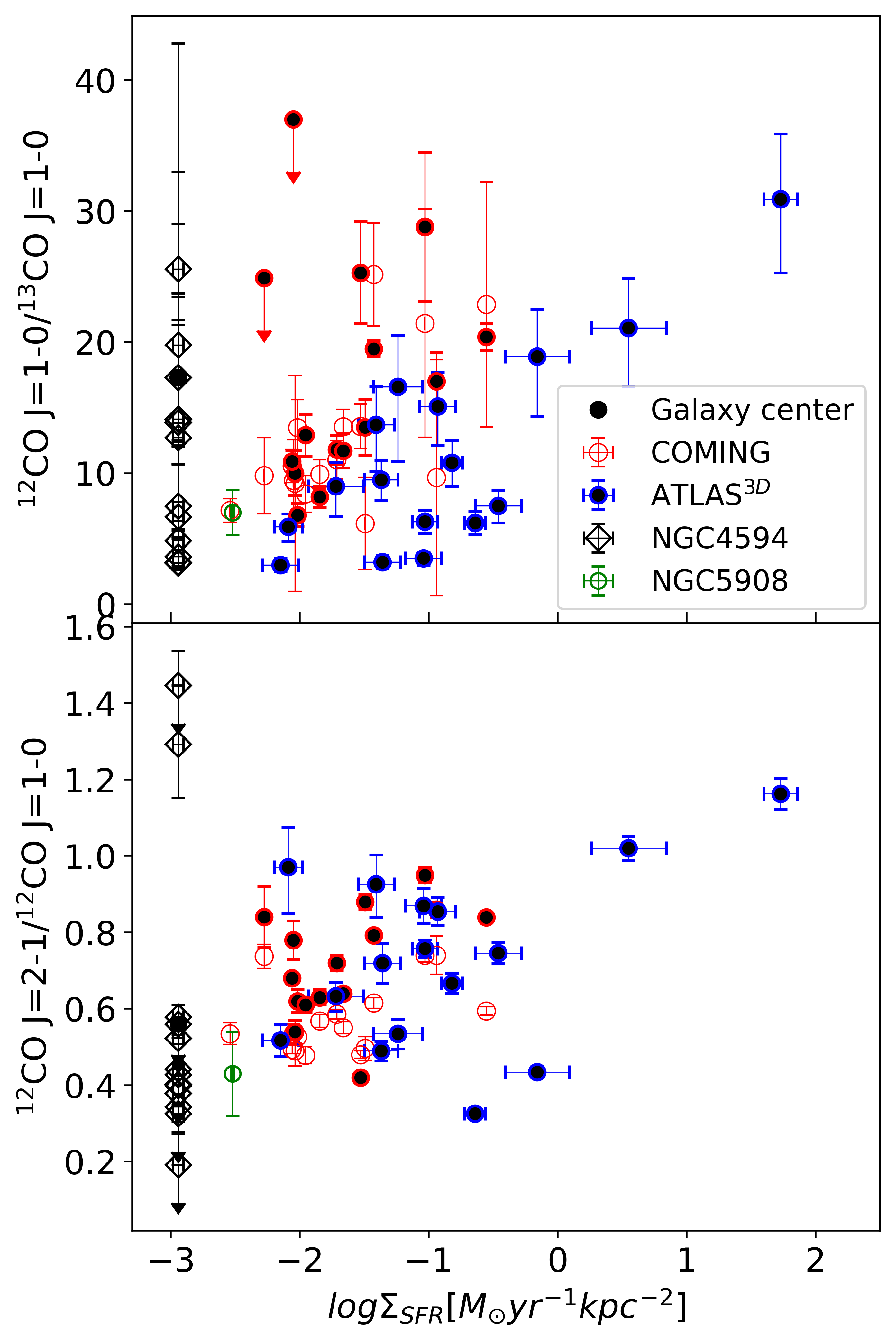, width=1.0\columnwidth}
\caption{CO line ratios at different SFR surface densities ($\Sigma_{\rm SFR}$). Symbols of different galaxy samples are the same as in Fig.~\ref{fig:lineratioandtypecode}.
}\label{fig:SFR&ratio}
\end{center}
\end{figure}

We further examine the dependence of the CO line ratios on the SFR surface density $\Sigma_{\rm SFR}$ of the galaxies. In Fig.~\ref{fig:SFR&ratio}, we compare the measurements of different positions of NGC~4594 to the COMING and ATLAS$\rm^{3D}$ samples. A two-dimensional Kolmogorove-Smirnov (KS) test (e.g., \citealt{Peacock83} \footnote[1]{https://github.com/syrte/ndtest}) of the COMING and ATLAS$\rm^{3D}$ samples on the $^{12}$CO/$^{13}$CO~$J=1-0-\Sigma_{\rm SFR}$ and $^{12}$CO~$J=2-1$/$J=1-0-\Sigma_{\rm SFR}$ diagrams results in a probability of $p=1.2\times 10^{-3}$ and $1.0\times 10^{-3}$, respectively. Here the $p$ value means the probability of the two samples from the same distribution. Typically we regard $p<0.05$ as the two samples being statistically different. Therefore, such low values of $p$ drawn on the $^{12}$CO/$^{13}$CO~$J=1-0-\Sigma_{\rm SFR}$ and $^{12}$CO~$J=2-1$/$J=1-0-\Sigma_{\rm SFR}$ diagrams indicate that the COMING and ATLAS$\rm^{3D}$ samples are significantly different in molecular gas properties. The different molecular gas properties in early- and late-type galaxies could be caused by both the metal enrichment and the optical depth effects, as discussed above. The two effects have similar impacts on the line ratios in different types of galaxies, and the combined effect is that a late-type galaxy with a generally more active SF in the near past is expected to have a higher $^{12}$CO/$^{13}$CO~$J=1-0$ ratio. Compared to most of the late-type galaxies from the COMING sample (as well as NGC~5908), the molecular gas in NGC~4594, except for two positions with extremely high $^{12}$CO~$J=2-1$/$J=1-0$, has systematically lower $^{12}$CO~$J=2-1$/$J=1-0$ line ratios. This indicates the gas is on average less heated by the young stars from SF regions. On the other hand, we do not find a significant dependence of the $^{12}$CO/$^{13}$CO~$J=1-0$ line ratio of the nuclear region of galaxies on $\Sigma_{\rm SFR}$ at $\Sigma_{\rm SFR}\lesssim1\rm~M_\odot~yr^{-1}~kpc^{-2}$ (the filled symbols in the upper panel of Fig.~\ref{fig:SFR&ratio}). This may indicate that the gas in the nuclear region is at least partially heated by other mechanisms instead of SF in these gas-poor and SF inactive galaxies.

\subsection{Star formation law}\label{subsec:SFR}

Our measurement of the molecular gas properties and the archival measurement of the IR properties of NGC~4594 indicate that it is an extremely gas poor [$M_{\rm H_2}/M_*\approx(1.1-1.5)\times10^{-3}$; Table~\ref{table:NGC4594}] and SF inactive galaxy (\S\ref{subsec:DiscussCOLineRatio}). In this section, we further examine how abnormal this inactivity is and what may be responsible for it. 

\begin{figure}
\begin{center}
\epsfig{figure=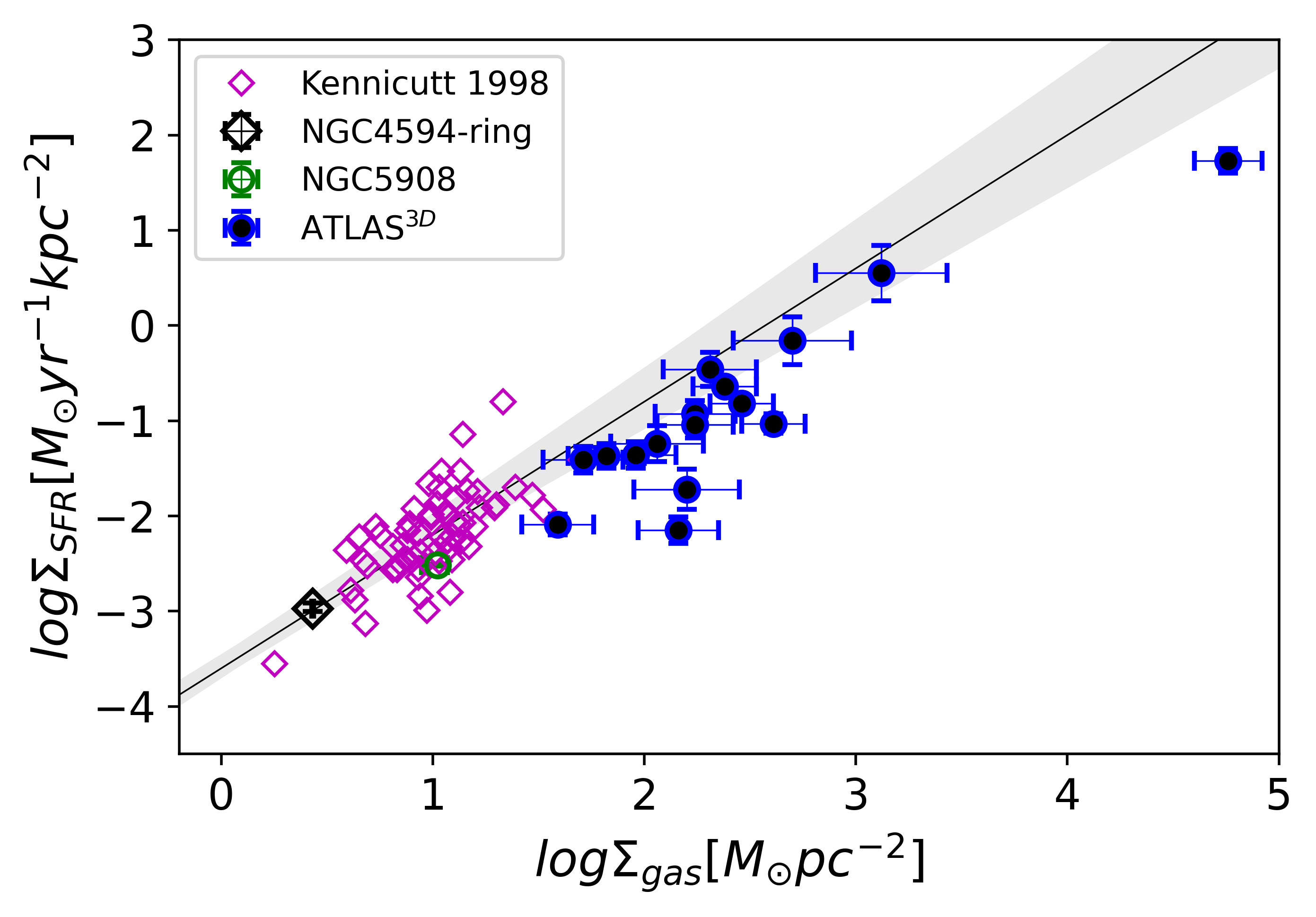, width=1.0\columnwidth}
\caption{Compare different galaxy samples on the star formation law. The solid line and the shaded area are the best-fit Kennicutt-Schmidt law and its 1~$\sigma$ uncertainty, while the purple diamonds are the galaxies used to define this relation, both from \citet{Kennicutt98}. Other symbols are the same as in Fig.~\ref{fig:lineratioandtypecode}. 
}\label{fig:SFlaw}
\end{center}
\end{figure}

In Fig.~\ref{fig:SFlaw}, we compare the selected galaxies from the ATLAS$\rm^{3D}$ sample (the COMING sample is not included in the plot as there is no uniform measurement of \ion{H}{1} listed in \citealt{Yajima21}) and \citet{Kennicutt98} to the two massive isolated spiral galaxies NGC~5908 and NGC~4594 on the Kennicutt-Schmidt SF law defined in \citet{Kennicutt98}: 
\begin{equation}\label{equi:SFR}
\frac{\sum_{\rm SFR}}{\rm M_{\odot}~yr^{-1}~kpc^{-2}}=(2.5\pm0.7)\times10^{-4} (\frac{\sum_{\rm gas}}{\rm M_{\odot}~pc^{-2}})^{1.4\pm0.15} ,
\end{equation}
where $\sum_{\rm SFR}$ is the disk-averaged SFR surface density, while $\sum_{\rm gas}$ is total atomic and molecular gas surface density. \citet{Bajaja84} shows that the radial distribution of the atomic gas is similar as that of the molecular gas in NGC~4594 (as shown in Fig.~\ref{fig:PVdiagram} or in the spatially resolved ALMA CO image as presented in \citealt{Sutter22}), both mainly on the dusty ring. Therefore, we compute the total atomic+molecular gas surface density $\sum_{\rm gas}$ of NGC~4594 by assuming a ring-like geometry of the gaseous ring with the inner and outer radius of 7.5 and 13.2~kpc. On the other hand, the SFR surface density $\sum_{\rm SFR}$ is directly obtained from \citet{Vargas19}, which is computed based on the spatially resolved IR and H$\alpha$ data. Cautions should be made that $\sum_{\rm gas}$ and $\sum_{\rm SFR}$ of the ATLAS$\rm^{3D}$ sample are measured only in the nuclear region of the galaxy, while those of other galaxies are measured across the entire disk.

As shown in Fig.~\ref{fig:SFlaw}, both NGC~5908 and NGC~4594 are well consistent with \citet{Kennicutt98}'s original sample which is used to define the SF law. This apparently indicates that there is no significant SF quenching in these massive isolated spiral galaxies, regardless of their gas content (NGC~5908 is more than an order of magnitude higher than NGC~4594 in cold gas content; \citealt{Li19}). The possibly slightly higher SF efficiency of NGC~4594 than NGC~5908 may be artificial due to the uncertainties in calculating their $\sum_{\rm gas}$ and $\sum_{\rm SFR}$. Furthermore, in an extremely gas-poor and massive galaxy like NGC~4594, additional heating sources other than the SF, such as UV photons from the evolved stars or weak AGN, etc., becomes important, which increase the IR and/or H$\alpha$ emissions used to calculate the SFR. \citet{Sutter22} found that the spatially resolved areas of NGC~4594 are mostly distributed below the Kennicutt-Schmidt law defined only with the molecular gas. Since they found poor correlation between the CO emission and the SF regions, the distribution of small areas extracted from NGC~4594 shows a large scatter on the Kennicutt-Schmidt law. Regardless of this large uncertainty, the suppress of SF in this massive early-type spiral galaxy is highly likely, although we do not find a significant inconsistency between NGC~4594 and the Kennicutt-Schmidt law defined with the total cold gas mass surface density in Fig.~\ref{fig:SFlaw}.

The early-type galaxies from the ATLAS$\rm^{3D}$ sample are systematically biased to the upper right corner in Fig.~\ref{fig:SFlaw}. This is partially caused by the measurement of $\sum_{\rm gas}$ and $\sum_{\rm SFR}$ only in the nuclear region, instead of averaged over the entire disk. These galaxies are also systematically below the Kennicutt-Schmidt law, indicating significant SF quenching probably via different mechanisms. For example, the galaxy with the highest $\sum_{\rm gas}$ and $\sum_{\rm SFR}$ (located in the most upper right of Fig.~\ref{fig:SFlaw}), NGC~1266, is known to have a strong molecular outflow driven by the AGN, which hinders the fragmentation and gravitational collapse necessary to form stars, and suppress the SF by a factor of $\gtrsim50$ \citep{Alatalo15b}. There may also be some other special SF quenching mechanisms work for massive early-type galaxies, such as morphological quenching caused by the gravitational stabilization of the star forming disk by the growth of a massive stellar spheroid (e.g., \citealt{Martig09}). However, just based on Fig.~\ref{fig:SFlaw}, we do not see any significant effect of these mechanisms in the representative massive spiral galaxy NGC~4594 or NGC~5908. 

\subsection{Gas circulation between NGC~4594 and its environment}\label{subsec:analysissamples}

Galaxies co-evolve with their environments largely via the circulation of the multi-phase gases in and out of the galactic disk and bulge. The multi-phase CGM could cool and be accreted onto the galactic disk, further condense and form stars, then be ejected out of the galaxies via stellar or AGN feedback. As one of the most massive isolated disk galaxies, NGC~4594 is unlikely strongly affected by major mergers in the near past, therefore provides us with an optimized laboratory to probe the steady state gas circulation in and out of galaxies.

We first roughly estimate the accretion rate of the multi-phase gases from the CGM. There are basically two types of gas accretion from the CGM, either via the radiative cooling then accretion of the hot CGM, or via the direct accretion of the cool atomic CGM. The radiative cooling rate of the hot CGM could be estimated from X-ray observations. Using the total mass of the hot CGM and its average radiative cooling timescale listed in \citet{Li13a}, we can estimate the average radiative cooling rate of the hot CGM around NGC~4594 to be $\dot{M}_{\rm hot}\sim0.06\rm~M_\odot~yr^{-1}$. This radiative cooling rate is estimated at $r\lesssim20\rm~kpc$ from the center of the galaxy, but since the density of the hot gas decreases fast toward larger radii, the very extended hot CGM does not cool as efficiently as in the inner region, so the total radiative cooling rate will not change a lot by adding the gas at larger radii. $\dot{M}_{\rm hot}$ estimated this way is about one order of magnitude lower than the observed current SFR (Table~\ref{table:NGC4594}). Therefore, even the current weak SF activity cannot be maintained only by the X-ray radiative cooling of the hot CGM. 

Another possible gas supply to maintain the SF activity is the direct accretion of cold atomic gas, via either a global radial gas inflow or some other forms with more complex gas kinematics. Direct measurements of the cold gas inflow rate is still rare, but in some galaxies, this cold mode accretion is indeed sufficient to replenish the gas consumed in SF (e.g., \citealt{Schmidt16, Di Teodoro21}). However, most of such galaxies with a strong gas inflow are gas-rich with $M_{\rm HI}$ typically at least one order of magnitude larger than that in NGC~4594 (Table~\ref{table:NGC4594}; e.g., \citealt{Schmidt16}). Furthermore, compared to the hot mode accretion which mostly dominates in massive galaxies such as NGC~4594, this code mode accretion often appears in low-mass galaxies and/or at high redshifts (e.g., \citealt{Keres05}). There is no reported detection of the extended \ion{H}{1} disk around NGC~4594. Existing observations only detect the \ion{H}{1} 21-cm line to an extent comparable to the dust ring or the area covered by our IRAM 30m observations of the CO lines (e.g., \citealt{Bajaja84}). Therefore, although we cannot rule out the possibility of the direct accretion of cold atomic gas as an important fuel to continue the SF, we herein also need to consider other possibilities.

In addition to the accretion of the external gas, another possible gas supply is the internal mass loss from evolved stars, dominated by those in the mass range of $0.8 \rm~M_\odot < M < 8~M_\odot$ in the format of planetary nebula (PN). Assuming the mass loss by each PN is $\sim0.3\rm~M_\odot$ \citep{Faber76} and a PN birth rate in our galaxy \citep{Peimbert93}, we obtain a total stellar mass loss rate of $\sim3\times10^{-12} \rm~M_\odot~yr^{-1}~L_\odot^{-1}$ from low and intermediate mass stars \citep{Li09}. Adopting the stellar mass of NGC~4594 (Table~\ref{table:NGC4594}), we obtain a total stellar mass loss rate of $\dot{M}_{\rm star}\sim0.9\rm~M_\odot~yr^{-1}$. We herein only consider the mass loss in the stage of PN, as the galaxy is early-type and very inactive in star formation, so the mass loss should be dominated by the collective contribution from low and intermediate mass stars. The estimated mass loss rate is apparently sufficient to compensate the gas consumed in SF, but there are two issues preventing it as a major gas source: the spatial distribution of the ejected mass from stars and the heating of the gas by Type~Ia SNe. The stellar mass loss follows the spatial distribution of low and intermediate mass stars, which is much more extended than the gas-rich and possibly SF active ring. On the other hand, the spatial distribution of the Type~Ia SNe is also expected to follow the spatial distribution of the stars, so the SNe could be well coupled with the ejected stellar mass and efficiently heat it. Adopting an empirical Type~Ia SNe birth rate of $4.4 \times 10^{-4} (M_*/10^{10} \rm M_\odot)~yr^{-1}$ \citep{Mannucci05} and an average SN energy release of $\sim10^{51}\rm~ergs$, we can obtain a specific energy per particle of $\sim4\rm~keV$ (assuming fully ionized pure Hydrogen gas). This is certainly an oversimplified estimate, but even if the thermalization efficiency is much lower than 100\%, the corresponding temperature is still too high as SF fuels. Such a high temperature of the stellar ejecta may be supported by the detection of metal-enriched gas in X-ray (e.g., \citealt{Li15}). Therefore, in a massive gas-poor galaxy with a huge bulge, such as NGC~4594, the stellar ejecta is more likely to be heated to an X-ray emitting temperature and entrained in a galactic outflow (e.g., \citealt{Li09,Tang09}), instead of cools, condenses and forms stars.

Although the above discussions on the gas sources of SF is largely uncertain, it is very likely that the galaxy cannot always maintain the current level of SF. Furthermore, if the galaxy mainly grew in isolation with the current SFR, the growth timescale will be $\sim650\rm~Gyr$, far larger than the Hubble time. Therefore, there must be a much more active starburst episode in the growth history of the galaxy, which produces most of the stellar mass observed in the present day. Galaxy archeology using the color and spatial distributions of globular clusters around NGC~4594 suggests that the galaxy has indeed experienced a major merger several Gyrs ago (e.g., \citealt{Cohen20}). This major merger may also explain the discovery of large-scale stellar streams around the galaxy (e.g., \citealt{Malin97,MartinezDelgado21}). However, a regular galactic disk as observed in NGC~4594 could unlikely survive through this merger triggered starburst episode(s), so it likely formed later at lower redshifts (e.g., \citealt{Mo98,Dekel09}). There could still be a significant amount of cool gas reservoir after the cease of the starburst, either from the leftover SF fuels or the accretion of the recycled gas from the CGM. This gas reservoir could be easily removed in a clustered environment via many environmental effects such as ram pressure stripping (e.g., \citealt{Bekki02,Goto03}). For an isolated galaxy like NGC~4594, the disk may gradually grew after the formation of the stellar bulge in a merger triggered starburst stage several Gyrs ago. In this gradual growth, the cold gas accumulated within the galaxy could also be removed via some internal processes, producing the gas-poor disk as observed today. For example, although Type~Ia SNe are not well coupled with the cool gas disk, the buoyant bubbles could still drag gas out of the cool gas disk, producing some vertical dusty filaments above the disk of some gas-poor galaxies (e.g., \citealt{Li09,Li15}). Furthermore, the SF efficiency may also be significantly suppressed due to morphological quenching from the huge stellar bulge or some other effects in gas-poor galaxies (e.g., \citealt{Martig09}). However, we do not see a significant decrease of the SF efficiency in NGC~4594 in response to these processes (\S\ref{subsec:SFR}). 

\section{Summary and Conclusion}\label{sec:summary}

We study the spatial distribution of molecular gas at the center and along the front side of the dust ring of the isolated massive spiral galaxy NGC~4594 with the IRAM 30M/EMIR. We firmly detect the $^{12}$CO~$J=1-0$ line at all 13 positions, while the $^{13}$CO~$J=1-0$ and $^{12}$CO~$J=2-1$ lines are only detected at about half of the positions (upper limits given at other positions). We calculate the ratios between different CO lines and other physical parameters such as the kinetic temperature and optical depth of the molecular gas. We also construct a position-velocity diagram based on the spatially resolved measurements of the $^{12}$CO~$J=1-0$ line. We further compare our measurements to other galaxy samples and discuss the implications. Below we summarize our key results and conclusions.

$\bullet$ The total molecular gas mass of NGC~4594 measured with the $^{12}$CO~$J=1-0$ and $^{13}$CO~$J=1-0$ line is $M_{\rm H_2} \approx 4\times10^{8}\rm~M_\odot$ and $M_{\rm H_2} \approx 3\times10^{8}\rm~M_\odot$, respectively. This corresponds to a gas-to-stellar mass ratio of $M_{\rm H_2}/M_*\sim10^{-3}$, indicating NGC~4594 is extremely gas poor.

$\bullet$ The median values of the $^{12}$CO/$^{13}$CO~$J=1-0$ and $^{12}$CO~$J=2-1$/$J=1-0$ line ratios, as well as the optical depth and kinetic temperature of the molecular clouds on the dust ring are $\approx 9.2$, 0.34, 0.12, and 41~K, respectively. Comparing with other galaxies with different gas contents and SF properties, these parameters indicate that NGC~4594 is also very inactive in SF, with no much additional heating of its molecular gas. 

$\bullet$ The position-velocity diagram of the molecular gas along the dust ring of NGC~4594 shows significant flattening at the outermost positions. With a simple hyperbolic tangent plus linear function, we measure the inclination corrected maximum rotation velocity of $V_{\rm max,corr}\approx381\rm~km~s^{-1}$, which is consistent with the archival measurements based on the \ion{H}{1} 21-cm line in the same radial range. Such a large rotation velocity suggests that NGC~4594 locates in a dark matter halo with a mass $M_{\rm200}\gtrsim10^{13}\rm~M_\odot$ --- one of the most massive isolated spiral galaxies in the local universe.

We compare the CO line ratios and other physical parameters measured at different positions of NGC~4594 to those of some selected galaxies from the COMING and ATLAS$\rm^{3D}$ samples. We find that the SF efficiency of NGC~4594 is consistent with that predicted by the best-fit Kennicutt-Schmidt law, so there is no evidence of enhanced SF quenching in this extremely massive, gas-poor, SF inactive spiral galaxy. The observed molecular gas could in principle be provided from a few external or internal sources, but none of them seem sufficient to replenish the cold gas consumed in SF. Therefore, it is most likely that the galaxy experienced a starburst stage at high redshift, forming the huge stellar bulge, then the recycled or leftover gas provides most of the fuels maintaining the gentle SF process at a rate close to what is observed today. This process may finally result in the formation of the galactic disk. More detailed galactic archaeology is needed to better understand the assembly history of these super-massive isolated spiral galaxies.

\bigskip
\noindent\textbf{\uppercase{acknowledgements}}
\smallskip\\
\noindent We are grateful to the National Science Foundation of China (grants No. E3GJ251110 and No. 11861131007) and the National Key Basic Research and Development Program of China (grant No. 2017YFA0402704) for their support of our research. We also would like to thank the staffs at IRAM, as well as Drs. Zhi-Yu Zhang, Rainer Beck, and Ralf-J\"{u}rgen Dettmar for their helps in the observations, data reductions, and scientific discussions of this project.

\bigskip
\noindent\textbf{\uppercase{Data Availability}}
\smallskip\\
\noindent The original IRAM 30m data could be shared by reasonable request to the corresponding author.

\clearpage


\begin{table*}
\caption{IRAM 30m Observation Log of NGC~4594.} 
\footnotesize
\hspace{-0.3in}
\begin{tabular}{cccccccccc}
\hline\hline
Position & RA,DEC & Date & $\tau_{\rm 225GHz}$ & Humidity & $n_{\rm obs}$ & $t_{\rm obs}$ & $t_{\rm exp,v}$ & $t_{\rm exp,h}$ & rms \\
& J2000 & & & & & min &min& min&mK \\
\hline
1 & 12:39:47.46,-11:37:29.92 & 10/17/2018 & 0.34 & 91.7\% & 1 & 39.2 & 39.2 & 39.2 & 3.7\\
& & 10/19/2018 & 0.46 & 69.85\% & 1\\
& & 1/29/2019 & 0.40,0.40 &69.6\%,67.4\% & 2\\
\hline
2 & 12:39:49.86,-11:37:32.00 & 10/17/2018 & 0.33 & 92.4\% & 1 & 19.6 & 19.6 & 19.6 & 8.3 \\
& & 10/19/2018 & 0.43 & 67.9\% & 1\\
\hline
3 & 12:39:52.26,-11:37:35.28 & 10/17/2018 & 0.36 & 92.4\% & 1 &39.2& 29.4 & 29.4 & 4.0 \\
& & 10/19/2018 & 0.45 & 67\% & 1\\
& & 1/29/2019 & 0.41,0.41 & 72.0\%,72.3\% & 2\\
\hline
4 & 12:39:54.66,-11:37:37.37 & 10/17/2018 & 0.28 & 91.7\% & 1 &39.2& 39.3&29.4 & 3.9 \\
& & 10/19/2018 & 0.24 & 72.7\% & 1\\
& & 1/29/2019 & 0.24,0.40 & 72.3\%,71.7\% & 2\\
\hline
5 & 12:39:57.05,-11:37:38.97 & 10/17/2018 & 0.33 & 88.2\% & 1 &39.2& 39.2&19.6 & 4.2 \\
& & 10/19/2018 & 0.50 & 76.1\% & 1\\
& & 1/29/2019 & 0.31,0.91 & 76.5\%,73.8\% & 2\\
\hline
6 & 12:39:59.45,-11:37:39.62 & 10/17/2018 & 0.30 & 93.2\% & 1 &39.2& 29.4&29.4 & 4.3\\
& & 10/19/2018 & 0.56 & 76.6\% & 1\\
& & 1/29/2019 & 0.34,0.35 & 78.1\%,78.4\% & 2\\
\hline
7 & 12:39:59.45,-11:37:22.84 & 10/19/2018 & 0.72 & 77.5\% & 1 &29.4& 19.6&29.4 & 4.4\\
& & 11/15/2018 & 0.55 & 89.7\% & 1\\
& & 1/29/2019 & 0.34 & 75.1\% & 1\\
\hline
8 & 12:40:1.04,-11:37:39.17 & 10/19/2018 & 0.50 & 76.2\% & 1 &29.4& 9.8&29.4 & 6.2\\
& & 11/15/2018 & 0.65 & 88.4\% & 1\\
& & 1/29/2019 & 0.30 & 73.8\% & 1\\
\hline
9 & 12:40:3.45,-11:37:37.66 & 10/19/2018 & 0.45 & 73.7\% & 1 &29.4& 19.6& 29.4 & 4.4\\
& & 11/15/2018 & 0.46 & 87.1\% & 1\\
& & 1/29/2019 & 0.30 & 74.3\% & 1\\
\hline
10 & 12:40:5.84,-11:37:35.18 & 10/19/2018 & 0.42 & 72.0\% & 1 &39.2& 19.6&19.6 & 5.6\\
& & 11/15/2018 & 0.59 & 87.2\% & 1\\
& & 1/29/2019 & 0.29,0.29 & 74.4\%,75.3\% & 2\\
\hline
11 & 12:40:8.24,-11:37:31.28& 10/19/2018 & 0.38 & 72.3\% & 1 &49.0& 29.4&39.2 & 3.3\\
& & 11/15/2018 & 0.66 & 85.4\% & 1\\
& & 11/16/2018 & 0.18 & 59.4\% & 1\\
& & 1/29/2019 & 0.28,0.34 & 76.1\%,75.5\% & 2\\
\hline
12 & 12:40:10.63,-11:37:27.13 & 10/19/2018 & 0.40 & 72.3\% & 1 &19.6& 9.8&19.6 & 7.2\\
& &11/16/2018 & 0.24 & 59.9\% & 1\\
\hline
13 & 12:40:13.03,-11:37:23.46 & 10/19/2018 & 0.45 & 72.3\% & 1 &49.0& 39.2&49.0 & 3.2\\
& & 11/16/2018 & 0.20,0.20 & 61.7\%,62.5\% & 2\\
& & 1/29/2019 & 0.33,0.32 & 78.6\%,83.4\% & 2\\
\hline\hline

\end{tabular}\label{table:NGC4594obslog}\\

Observations taken in October and November 2018 are from project 063-18, while those taken in January 2019 are from project 189-18. The opacity at 225~GHz ($\tau_{\rm 225GHz}$) and the humidity are the average values during each observation scan. $n_{\rm obs}$ is the number of observation scans at each position. Each scan has an effective on-source exposure time of 9.8~min. $t_{\rm obs}$ is the total on-source exposure time at each position. $t_{\rm exp,v}$ and $t_{\rm exp,h}$ are the total effective exposure time of the vertical and horizontal polarization components of the $^{12}$CO~$J=1-0$ spectra, after filtering the bad data. The rms is calculated based on fitting the baseline of the combined $^{12}$CO~$J=1-0$ spectra of both polarization components.

\end{table*}


\begin{table*}
\begin{center}
\rotatebox{90}{
\begin{minipage}{\textheight}
\caption{Observed and Derived Parameters of the CO Lines}
\setlength{\leftskip}{-40pt}
\footnotesize
\tabcolsep=3.0pt
\hspace{-1in}
\begin{tabular}{cccccccccccccc}
\hline
Region & $d$ & $I_{\rm ^{12}CO_{\rm 10}}$ & $v_{\rm ^{12}CO_{\rm 10}}$ & $I_{\rm ^{13}CO_{\rm 10}}$ & $v_{\rm ^{13}CO_{\rm 10}}$ & $I_{\rm ^{12}CO_{\rm 21}}$ & $v_{\rm ^{12}CO_{\rm 21}}$ & ${\rm \mathsmaller{\mathsmaller{\frac{^{12}CO_{\rm 10}}{^{13}CO_{\rm 10}}}}}$ & ${\rm \mathsmaller{\mathsmaller{\frac{^{12}CO_{\rm 21}}{^{12}CO_{\rm 10}}}}}$ & $N_{\rm H_2,^{12}CO}$ & $N_{\rm H_2,^{13}CO}$ & $\tau(\rm ^{13}CO)$ & $T_{\rm K}$ \\
 & $^{\rm kpc}$ & $\rm K~km~s^{-1}$ & $\rm km~s^{-1}$ & $\rm K~km~s^{-1}$ & $\rm km~s^{-1}$ & $\rm K~km~s^{-1}$ & $\rm km~s^{-1}$ & & & $f^{-1}10^{21}\rm~cm^{-2}$ & $f^{-1}10^{21}\rm~cm^{-2}$ &  & K \\
\hline

1 & 11.09 & $ 1.25\pm 0.11 $ & $ -290.4\pm 1.6 $ & $ 0.31\pm 0.06 $ & $ -302.9\pm 5.0 $ & $ \textless~0.90 $ & $ - $ & $ 3.18~^{+1.96}_{-0.44} $ & $ \textless~0.33 $& $ 0.25\pm 0.11 $ & $ 0.31\pm 0.06 $ & $ 0.37~^{+0.37}_{-0.43} $ & $ 11.66~^{+10.47}_{-1.82} $\\
2 & 8.87 & $ 3.50\pm 0.25 $ & $ -279.9\pm 0.1 $ & $ 0.42\pm 0.11 $ & $ -285.7\pm 0.1 $ & $ 2.52\pm 0.48 $ & $ -290.0\pm 0.2 $ & $ 6.68~^{+4.01}_{-1.05} $ & $ 0.34~^{+0.07}_{-0.06} $& $ 0.70\pm 0.25 $ & $ 0.42\pm 0.11 $ & $ 0.16~^{+0.16}_{-0.13} $ & $ 25.99~^{+20.05}_{-5.44} $\\
3 & 6.65 & $ 1.76\pm 0.15 $ & $ -197.2\pm 3.5 $ & $ 	\textless~0.38 $ & $ - $ & $ 4.84\pm 0.59 $ & $ -192.7\pm 5.5 $ & $ 19.79~^{+13.19}_{-2.78} $ & $ 	1.29~^{+0.24}_{-0.14} $& $ 0.35\pm 0.15 $ & $ 	\textless~0.38 $ & $ 	\textless~0.23 $ & $ 69.02~^{+61.26}_{-17.71} $\\
4 & 4.43 & $ 1.89\pm 0.16 $ & $ -135.1\pm 4.5 $ & $ 	\textless~0.31 $ & $ - $ & $ 1.72\pm 0.30 $ & $ -123.5\pm 4.1 $ & $ 25.59~^{+17.19}_{-3.88} $ & $ 	0.44~^{+0.08}_{-0.09} $& $ 0.38\pm 0.16 $ & $ 	\textless~0.31 $ & $ 	\textless~0.16 $ & $ 83.98~^{+72.77}_{-17.08} $\\
5 & 2.22 & $ 1.82\pm 0.16 $ & $ -34.5\pm 4.1 $ & $ 0.19\pm 0.06 $ & $ 0.4\pm 4.1 $ & $ 2.01\pm 0.30 $ & $ -40.5\pm 6.5 $ & $ 7.48~^{+4.51}_{-1.13} $ & $ 0.52~^{+0.09}_{-0.10} $& $ 0.36\pm 0.16 $ & $ 0.19\pm 0.06 $ & $ 0.14~^{+0.15}_{-0.09} $ & $ 29.54~^{+22.40}_{-6.19} $\\
6 & 0.00 & $ 1.76\pm 0.17 $ & $ 60.8\pm 3.4 $ & $ 	\textless~0.55 $ & $ - $ & $ 1.34\pm 0.29 $ & $ 55.8\pm 5.3 $ & $ 13.86~^{+9.61}_{-1.49} $ & $ 	0.40~^{+0.11}_{-0.08} $& $ 0.35\pm 0.17 $ & $ 	\textless~0.55 $ & $ 	\textless~0.32 $ & $ 50.31~^{+46.37}_{-9.06} $\\
7 & 0.00 & $ 1.05\pm 0.12 $ & $ 75.9\pm 3.0 $ & $ 	\textless~0.27 $ & $ - $ & $ 	\textless~1.16 $ & $ - $ & $ 17.31~^{+11.73}_{-3.48} $ & $ 	\textless~0.56 $& $ 0.21\pm 0.12 $ & $ 	\textless~0.27 $ & $ 	\textless~0.22 $ & $ 59.48~^{+54.49}_{-16.66} $\\
8 & -1.48 & $ 1.94\pm 0.22 $ & $ 122.6\pm 4.7 $ & $ 0.32\pm 0.07 $ & $ 139.5\pm 4.8 $ & $ 1.55\pm 0.26 $ & $ 114.6\pm 3.0 $ & $ 4.86~^{+2.95}_{-0.71} $ & $ 0.38~^{+0.08}_{-0.07} $& $ 0.39\pm 0.22 $ & $ 0.32\pm 0.07 $ & $ 0.23~^{+0.24}_{-0.18} $ & $ 17.98~^{+14.87}_{-4.15} $\\
9 & -3.70 & $ 1.40\pm 0.14 $ & $ 201.8\pm 2.5 $ & $ 	\textless~0.42 $ & $ - $ & $ 1.16\pm 0.35 $ & $ 199.3\pm 5.2 $ & $ 14.11~^{+9.59}_{-1.61} $ & $ 	0.40\pm 0.13 $& $ 0.28\pm 0.14 $ & $ 	\textless~0.42 $ & $ 	\textless~0.31 $ & $ 51.83~^{+47.53}_{-8.13} $\\
10 & -5.92 & $ 1.00\pm 0.17 $ & $ 297.6\pm 4.2 $ & $ 	\textless~0.33 $ & $ - $ & $ 	\textless~3.21 $ & $ - $ & $ 12.71~^{+8.63}_{-2.02} $ & $ 	\textless~1.45 $& $ 0.20\pm 0.17 $ & $ 	\textless~0.33 $ & $ 	\textless~0.33 $ & $ 46.81~^{+42.91}_{-9.58} $\\
11 & -8.13 & $ 3.16\pm 0.14 $ & $ 398.0\pm 0.1 $ & $ 0.79\pm 0.09 $ & $ 376.3\pm 0.1 $ & $ \textless~1.34 $ & $ - $ & $ 3.15~^{+1.87}_{-0.56} $ & $ \textless~0.19 $& $ 0.63\pm 0.14 $ & $ 0.79\pm 0.09 $ & $ 0.38~^{+0.43}_{-0.27} $ & $ 11.45~^{+10.27}_{-1.73} $\\
12 & -10.35 & $ 2.46\pm 0.19 $ & $ 427.6\pm 1.7 $ & $ 0.54\pm 0.11 $ & $ 419.2\pm 5.0 $ & $ \textless~2.26 $ & $ - $ & $ 3.63~^{+2.14}_{-0.72} $ & $ \textless~0.43 $& $ 0.49\pm 0.19 $ & $ 0.54\pm 0.11 $ & $ 0.32~^{+0.35}_{-0.24} $ & $ 13.06~^{+11.28}_{-3.80} $\\
13 & -12.57 & $ 0.72\pm 0.12 $ & $ 417.6\pm 5.0 $ & $ 	\textless~0.21 $ & $ - $ & $ 	\textless~1.03 $ & $ - $ & $ 14.14~^{+9.62}_{-2.02} $ & $ 	\textless~0.58 $& $ 0.14\pm 0.12 $ & $ 	\textless~0.21 $ & $ 	\textless~0.31 $ & $ 51.70~^{+47.71}_{-11.22} $\\

\hline
\end{tabular}\label{table:COlinepara}\\\\
Region labels of different observations are plotted in Fig.~\ref{fig:IRAMbeam}. $d$ is the distance to the minor axis of the galaxy. $I$ is the integrated line intensity of different lines, which has been corrected for the main beam and forward efficiencies. If the lines are not firmly detected, we list their $3\sigma$ upper limit, while fill ``-" in the velocity column. The line ratios are corrected for beam dilution. $N_{\rm H_2,^{12}CO}$ and $N_{\rm H_2,^{13}CO}$ are the molecular gas column density derived from $I_{\rm ^{12}CO~J=1-0}$ and $I_{\rm ^{13}CO~J=1-0}$, respectively. $N_{\rm H_2,^{12}CO}$ and $N_{\rm H_2,^{13}CO}$ depend on the filling factor $f$ within the beam as $\propto f^{-1}$. $\tau(\rm ^{13}CO)$ and $T_{\rm K}$ are the optical depth and the kinetic temperature.
\end{minipage}
}
\end{center}
\end{table*}


\begin{thebibliography}{34}
\expandafter\ifx\csname natexlab\endcsname\relax\def\natexlab#1{#1}\fi

\bibitem[Alatalo et al.(2015)]{Alatalo15a} Alatalo, K., Crocker, A.~F., Aalto, S., et al.\ 2015a, \mnras, 450, 3874.

\bibitem[Alatalo et al.(2015)]{Alatalo15b} Alatalo, K., Lacy, M., Lanz, L., et al.\ 2015b, \apj, 798, 31. 

\bibitem[Bajaja et al.(1984)]{Bajaja84} Bajaja, E., van der Burg, G., Faber, S.~M., et al.\ 1984, \aap, 141, 309.

\bibitem[Bajaja et al.(1991)]{Bajaja91} Bajaja, E., Krause, M., Dettmar, R.~J., et al.\ 1991, \aap, 241, 411.

\bibitem[Bekki et al.(2002)]{Bekki02} Bekki, K., Couch, W.~J., Shioya, Y., 2002, ApJ, 577, 651. 

\bibitem[Bolatto et al.(2013)]{Bolatto13} Bolatto A. D., Wolfire M., Leroy A. K.,\ 2013, ARA\&A, 51, 207.

\bibitem[Braine et al.(1993)]{Braine93} Braine, J., Combes, F., Casoli, F., et al.\ 1993, \aaps, 97, 887.

\bibitem[Calzetti et al.(2007)]{Calzetti07} Calzetti, D., Kennicutt, R.~C., Engelbracht, C.~W., et al.\ 2007, \apj, 666, 870. 

\bibitem[Carter et al.(2012)]{Carter12} Carter, M., Lazareff, B., Maier1, D., Chenu1, J.-Y., Fontana, A.-L., et al., 2012, A\&A 538, 89.

\bibitem[Cohen et al.(2020)]{Cohen20} Cohen, R.~E., Goudfrooij, P., Correnti, M., et al.\ 2020, \apj, 890, 52. 

\bibitem[Cormier et al.(2018)]{Cormier2018} Cormier, D., Bigiel, F., Jim{\'e}nez-Donaire, M.~J., et al.\ 2018, \mnras, 475, 3909.

\bibitem[Crocker et al.(2012)]{Crocker12} Crocker, A., Krips, M., Bureau, M., et al.\ 2012, \mnras, 421, 1298. 

\bibitem[Davis et al.(2014)]{Davis14} Davis, T.~A., Young, L.~M., Crocker, A.~F., et al.\ 2014, \mnras, 444, 3427. 

\bibitem[Dekel et al.(2009)]{Dekel09} Dekel, A., Sari, R., Ceverino, D., 2009, \apj, 703, 785. 

\bibitem[Di Teodoro \& Peek(2021)]{Di Teodoro21} Di Teodoro, E.~M. \& Peek, J.~E.~G.\ 2021, \apj, 923, 220.

\bibitem[Faber \& Gallagher(1976)]{Faber76} Faber, S.~M. \& Gallagher, J.~S.\ 1976, \apj, 204, 365.

\bibitem[Frerking et al.(1982)]{Frerking82} Frerking, M.~A., Langer, W.~D., \& Wilson, R.~W.\ 1982, \apj, 262, 590.

\bibitem[Gao(1996)]{Gao96}Gao, Y. 1996, PhD thesis, State Univ. of New York.

\bibitem[Gao \& Solomon(2004)]{Gao04} Gao, Y. \& Solomon, P.~M.\ 2004, \apj, 606, 271. 

\bibitem[Goto et al.(2003)]{Goto03} Goto, T., Okamura, S., Sekiguchi, M., et al., 2003, \pasj, 55, 757. 

\bibitem[Hasegawa(1997)]{Hasegawa97} Hasegawa T., 1997, IAUS, 170, 39.

\bibitem[Heald et al.(2022)]{Heald21} Heald, G.~H., Heesen, V., Sridhar, S.~S., et al.\ 2022, \mnras, 509, 658.

\bibitem[Irwin et al.(2012a)]{Irwin12a} Irwin, J., Beck, R., Benjamin, R.~A., et al.\ 2012a, \aj, 144, 43.

\bibitem[Irwin et al.(2012b)]{Irwin12b} Irwin, J., Beck, R., Benjamin, R.~A., et al.\ 2012b, \aj, 144, 44.

\bibitem[Irwin et al.(2017)]{Irwin17} Irwin, J. A., Schmidt, P., Damas-Segovia, A., et al., 2017, MNRAS, 464, 1333.

\bibitem[Irwin et al.(2019a)]{Irwin19a} Irwin, J. A., Damas-Segovia, A., Krause, M., et al., 2019a, Galaxies, 7, 42.

\bibitem[Irwin et al.(2019b)]{Irwin19b} Irwin, J. A., Wiegert, T., Merritt, A., et al., 2019b, AJ, 158, 21.

\bibitem[Jardel et al.(2011)]{Jarrett11} Jardel, J.~R., Gebhardt, K., Shen, J., et al.\ 2011, \apj, 739, 21.

\bibitem[Jarrett et al.(2019)]{Jarrett19} Jarrett, T.~H., Cluver, M.~E., Brown, M.~J.~I., et al.\ 2019, \apjs, 245, 25.

\bibitem[Jim{\'e}nez-Donaire et al.(2017)]{Jim17} Jim{\'e}nez-Donaire, M.~J., Cormier, D., Bigiel, F., et al.\ 2017, \apjl, 836, L29.

\bibitem[Kennicutt(1998)]{Kennicutt98} Kennicutt R. C., 1998, ApJ, 498, 541.

\bibitem[Kennicutt et al.(2003)]{Kennicutt03} Kennicutt, R.~C., Armus, L., Bendo, G., et al.\ 2003, \pasp, 115, 928. 

\bibitem[Kennicutt \& Evans(2012)]{Kennicutt12} Kennicutt R. C., Evans N. J., 2012, ARA\&A, 50, 531.

\bibitem[Kere{\v{s}} et al.(2005)]{Keres05} Kere{\v{s}}, D., Katz, N., Weinberg, D.~H., et al.\ 2005, \mnras, 363, 2. 

\bibitem[Krause et al.(2018)]{Krause18} Krause, M., Irwin, J., Wiegert, T., et al., 2018, A\&A, 611, 72.

\bibitem[Krause et al.(2020)]{Krause20} Krause, M., Irwin, J., Schmidt, P., et al., 2020, A\&A, 639, A112.

\bibitem[Leroy et al.(2022)]{Leroy22} Leroy, A.~K., Rosolowsky, E., Usero, A., et al.\ 2022, \apj, 927, 149.

\bibitem[Li et al.(2009)]{Li09} Li, J.-T., Wang, Q.~D., Li, Z., et al., 2009, ApJ, 706, 693 

\bibitem[Li \& Wang(2013a)]{Li13a} Li, J.-T. \& Wang, Q.~D.\ 2013a, \mnras, 428, 2085.

\bibitem[Li \& Wang(2013b)]{Li13b} Li, J.-T. \& Wang, Q.~D.\ 2013b, \mnras, 435, 3071.

\bibitem[Li et al.(2014)]{Li14} Li, J.-T., Wang, Q. D., Crain, R. A., 2014, \mnras, 440, 859.

\bibitem[Li(2015)]{Li15} Li, J.-T., 2015, MNRAS, 453, 1062. 

\bibitem[Li et al.(2016)]{Li16} Li, J.-T., Beck, R., Dettmar, R.-J., et al.\ 2016, \mnras, 456, 1723. 

\bibitem[Li et al.(2017)]{Li17} Li, J.-T., Bregman, J.~N., Wang, Q.~D., et al.\ 2017, \apjs, 233, 20.

\bibitem[Li et al.(2019)]{Li19} Li, J.-T., Zhou, P., Jiang, X., et al.\ 2019, \apj, 877, 3. 

\bibitem[Lisenfeld et al.(2019)]{Ute19} Lisenfeld, U., Xu, C.~K., Gao, Y., et al.\ 2019, \aap, 627, A107. 

\bibitem[Liszt(2017)]{Liszt17} Liszt, H.~S.\ 2017, \apj, 835, 138.

\bibitem[Lu et al.(2023)]{Lu23} Lu, L.-Y., Li, J.-T., Vargas, C.~J., et al.\ 2023, \mnras, 519, 6098.

\bibitem[Mannucci et al.(2005)]{Mannucci05} Mannucci, F., Della Valle, M., Panagia, N., et al.\ 2005, \aap, 433, 807.

\bibitem[Malin \& Hadley(1997)]{Malin97} Malin, D., Hadley, B., 1997, Publ. Astron. Soc. Australia, 14, 52. 

\bibitem[Martig et al.(2009)]{Martig09} Martig, M., Bournaud, F., Teyssier, R., et al., 2009, \apj, 707, 250. 

\bibitem[Mart{\'\i}nez-Delgado et al.(2021)]{MartinezDelgado21} Mart{\'\i}nez-Delgado, D., Rom{\'a}n, J., Erkal, D., et al., 2021, MNRAS, 506, 5030.

\bibitem[Mo et al.(1998)]{Mo98} Mo, H.~J., Mao, S., White, S.~D.~M., 1998, \mnras, 295, 319. 

\bibitem[Muraoka et al.(2016)]{Muraoka16} Muraoka, K., Sorai, K., Kuno, N., et al.\ 2016, \pasj, 68, 89.

\bibitem[Ogle et al.(2016)]{Ogle16} Ogle P. M., Lanz L., Nader C., Helou G., 2016, ApJ, 817, 109. 

\bibitem[Peacock(1983)]{Peacock83} Peacock, J.~A.\ 1983, \mnras, 202, 615. 

\bibitem[Peimbert(1993)]{Peimbert93} Peimbert, M.\ 1993, Planetary Nebulae, 155, 523.

\bibitem[Pe{\~n}aloza et al.(2017)]{Penaloza17} Pe{\~n}aloza, C.~H., Clark, P.~C., Glover, S.~C.~O., et al.\ 2017, \mnras, 465, 2277.

\bibitem[Perlman et al.(2021)]{Perlman21} Perlman, E., Meyer, E. T., Wang, Q. D., et al., 2021, submitted to ApJ.

\bibitem[Sage et al.(1991)]{Sage91} Sage, L.~J., Mauersberger, R., \& Henkel, C.\ 1991, \aap, 249, 31.

\bibitem[Salucci et al.(2007)]{Salucci07} Salucci, P., Lapi, A., Tonini, C., et al.\ 2007, \mnras, 378, 41. 

\bibitem[Schmidt et al.(2016)]{Schmidt16} Schmidt, T.~M., Bigiel, F., Klessen, R.~S., et al.\ 2016, \mnras, 457, 2642.

\bibitem[Shetty et al.(2011)]{Shetty11a} Shetty, R., Glover, S.~C., Dullemond, C.~P., et al.\ 2011, \mnras, 412, 1686. 

\bibitem[Shetty et al.(2011)]{Shetty11b} Shetty, R., Glover, S.~C., Dullemond, C.~P., et al.\ 2011, \mnras, 415, 3253. 

\bibitem[Stein et al.(2022)]{Stein22} Stein, M., Heesen, V., Dettmar, R.-J., et al.\ 2022, arXiv:2210.07709

\bibitem[Sutter \& Fadda(2022)]{Sutter22} Sutter, J. \& Fadda, D.\ 2022, arXiv:2210.13527.

\bibitem[Tan et al.(2011)]{Tan11} Tan Q.-H., Gao Y., Zhang Z.-Y., Xia X.-Y., 2011, RAA, 11, 787.

\bibitem[Tang et al.(2009)]{Tang09} Tang, S., Wang, Q.~D., Lu, Y., et al.\ 2009, \mnras, 392, 77. 

\bibitem[Vargas et al.(2018)]{Vargas18} Vargas, C.~J., Mora-Partiarroyo, S.~C., Schmidt, P., et al.\ 2018, \apj, 853, 128. 

\bibitem[Vargas et al.(2019)]{Vargas19} Vargas, C.~J., Walterbos, R.~A.~M., Rand, R.~J., et al.\ 2019, \apj, 881, 26. 

\bibitem[Veilleux et al.(2009)]{Veilleux09} Veilleux, S., Rupke, D. S. N., Swaters, R., 2009, ApJL, 700, 149.

\bibitem[Wiegert et al.(2015)]{Wiegert15} Wiegert, T., Irwin, J., Miskolczi, A., et al.\ 2015, \aj, 150, 81. 

\bibitem[Wu et al.(2010)]{Wu10} Wu, J., Evans, N.~J., Shirley, Y.~L., et al.\ 2010, \apjs, 188, 313. 

\bibitem[Yajima et al.(2021)]{Yajima21} Yajima, Y., Sorai, K., Miyamoto, Y., et al.\ 2021, \pasj, 73, 257. 

\bibitem[Yang et al., in prep.]{Yang23} Yang Y., et al., in prep.

\bibitem[Yoon et al.(2021)]{Yoon21} Yoon, Y., Park, C., Chung, H., et al.\ 2021, \apj, 922, 249.

\bibitem[Young \& Scoville(1991)]{Young91} Young, J.~S. \& Scoville, N.~Z.\ 1991, \araa, 29, 581. 

\bibitem[Young et al.(2011)]{Young11} Young, L.~M., Bureau, M., Davis, T.~A., et al.\ 2011, \mnras, 414, 940.

\bibitem[Zheng et al.(2022a)]{Zheng22a} Zheng, Y., Wang, J., Irwin, J., et al.\ 2022a, \mnras, 513, 1329. 

\bibitem[Zheng et al.(2022b)]{Zheng22b} Zheng, Y., Wang, J., Irwin, J., et al.\ 2022b, Research in Astronomy and Astrophysics, 22, 085004. 


\end{thebibliography}
\end{document}